\documentclass[11pt]{article}
\usepackage[dvips]{graphics}
\input{amssym.def}
\input{amssym.tex}

\title{\Large\bf DIFFRACTION  OF AN OBLIQUELY INCIDENT
ELECTROMAGNETIC WAVE BY AN IMPEDANCE RIGHT-ANGLED CONCAVE WEDGE}
\author{Y.A. ANTIPOV\\
antipov@math.lsu.edu\\ 
Department of Mathematics, Louisiana State University\\
Baton Rouge LA 70803, USA}

\newcommand{\beqa}{\begin{eqnarray}}
\newcommand{\eeqa}[1]{\label{#1}\end{eqnarray}}
\newcommand{\bequ}{\begin{equation}}
\newcommand{\eequ}[1]{\label{#1}\end{equation}}

\newcommand{\Md}{\partial}

\newcommand{\ov}[1]{\overline{#1}}

\newcommand{\Ga}{\alpha}
\newcommand{\Gb}{\beta}
\newcommand{\Gd}{\delta}
\newcommand{\Ge}{\epsilon}
\newcommand{\Gve}{\varepsilon}
\newcommand{\Gf}{\phi}
\newcommand{\Gvf}{\varphi}
\newcommand{\Gg}{\gamma}
\newcommand{\Gc}{\chi}

\newcommand{\Gk}{\kappa}

\newcommand{\Gl}{\lambda}
\newcommand{\Gn}{\eta}

\newcommand{\Gt}{\theta}

\newcommand{\Gr}{\rho}

\newcommand{\Gs}{\sigma}

\newcommand{\Go}{\omega}
\newcommand{\Gx}{\xi}
\newcommand{\Gy}{\psi}
\newcommand{\Gz}{\zeta}
\newcommand{\GD}{\Delta}
\newcommand{\GF}{\Phi}
\newcommand{\GG}{\Gamma}
\newcommand{\GL}{\Lambda}

\newcommand{\CC}{{\cal C}}

\newcommand{\CE}{{\cal E}}

\newcommand{\CG}{{\cal G}}
\newcommand{\CH}{{\cal H}}

\newcommand{\CL}{{\cal L}}

\newcommand{\CT}{{\cal T}}

\def\Ba{{\bf a}}
\def\Bb{{\bf b}}

\def\BK{{\bf K}}

\newcommand{\beq}{\begin{equation}}
\newcommand{\eeq}{\end{equation}}
\newcommand{\barr}{\begin{eqnarray}}
\newcommand{\earr}{\end{eqnarray}}
\newcommand{\beqn}{\begin{equation*}}
\newcommand{\eeqn}{\end{equation*}}
\newcommand{\barrn}{\begin{eqnarray*}}
\newcommand{\earrn}{\end{eqnarray*}}

\newcommand{\fr}{\frac}

\newcommand{\diag}{\mbox{diag}}

\newcommand{\sgn}{\mathop{\rm sgn}\nolimits}

\newcommand{\ind}{\mathop{\rm ind}\nolimits}
\newcommand{\I}{\mathop{\rm Im}\nolimits}
\newcommand{\R}{\mathop{\rm Re}\nolimits}
\newcommand{\const}{\mbox{const}}

\begin{document}
\maketitle

\begin{abstract}

Scattering of a plane electromagnetic  wave by an anisotropic impedance right-angled  concave wedge at skew incidence is analyzed.
A closed-form solution is derived by reducing the problem  to a symmetric order-2 vector Riemann-Hilbert problem (RHP)
on the real axis. The problem of matrix factorization  leads to a scalar  RHP
on a genus-3 Riemann surface. Its solution is derived by the Weierstrass integrals.
Due to a special symmetry of the problem the associated Jacobi
inversion problem is solved in terms of elliptic integrals, not a  genus-3 Riemann $\Gt$-function.  
The electric and magnetic field components
are expressed through the Sommerfeld integrals, and the incident and reflected
waves are recovered. 
\end{abstract}

\setcounter{equation}{0}

\section{Introduction}

In diffraction theory the study of scattering of a plane obliquely incident electromagnetic
wave from an anisotropic impedance wedge is one of the key canonical problems.
In the case of normal incidence when the tensor impedance has zero diagonal and nonzero off-diagonal
entries,   the problem was solved in {\bf(\ref{mal1})}. For oblique incidence,
a closed-form solution is known for some special cases including the vector case for
a half-plane 
{\bf(\ref{hur})}, {\bf(\ref{lue})}, {\bf(\ref{ant1})} and  the diagonal  and  triangular cases for a wedge when the impedance parameters meet certain 
conditions allowing for reduction of the problem to separately and  consequently solved  scalar  Maliuzhinets equations
{\bf(\ref{ber})}, {\bf(\ref{lya1})}. 

Approximate solutions by the perturbation technique are available for the cases when the wedge is almost a
half-plane {\bf(\ref{sye})} or when the incidence is almost normal {\bf(\ref{pel})}.
Different approximate numerical solutions are available for the general case. They include those obtained by the method
of integral equations 
{\bf(\ref{lya2})}, {\bf(\ref{ant2})}, {\bf(\ref{lya})}, the method of approximate matrix factorization in conjunction with the Fredholm integral equation theory  {\bf(\ref{dan1})}, 
 and the probabilistic random walk method {\bf(\ref{bud})}.  
 
A scalar diffraction problem for a concave wedge was analyzed in {\bf(\ref{sha})} by means of a
 Carleman-type 
boundary value problem of the theory of analytic functions.
A method based on
splitting the spectra into two functions, the solution of the  Maliuzhinets equation
 and a function that
is defined by a series whose coefficients solve a certain 
infinite system of linear algebraic equations, was recently proposed in {\bf(\ref{osi1})}.
By this method, the diffraction field in the case
of scattering by a wedge of angle $3\pi/2$
was recovered in  {\bf(\ref{osi2})}.
A method of reduction of wedge diffraction problems to functional equations was presented in {\bf(\ref{dan2})}.
In particular, it was pointed out that the electromagnetic problem for a right-angled convex wedge
could be reduced to an order-6 vector RHP. 

An exact solution for the general case of the tensor impedance and obliquely incident electromagnetic wave
even for a right-angled 
wedge is still unavailable in the literature. Our goal in this paper was to derive a closed-form solution
of the problem on 
an impenetrable right-angled concave  wedge (of angle $3\pi/2$) in the case of oblique incidence
when on the faces $S^\pm$ of the wedge the boundary conditions
are  
\beq
\pm\CE_z=\Gn_{\Gr\Gr}^\pm Z\CH_\Gr,\quad   \mp\CE_\Gr=\Gn_{z z}^\pm Z \CH_z, \quad (\Gr,\Gt)\in S^\pm.
\label{1.1}
\eeq
Here,  $S^-=\{\Gt=0, 0<\Gr<\infty\}$, $S^+=\{\Gt=\pi/2, 0<\Gr<\infty\}$,
$\CE_\Gr, \CH_\Gr$
and $\CE_z, \CH_z$
are the $\Gr$- and $z$-components of the electromagnetic field,  $Z$ is 
 the intrinsic 
impedance of the medium, and $\Gn_{\Gr\Gr}^\pm$ and 
$\Gn_{zz}^\pm$ are the impedance parameters. 

In Section 2, we constitute the boundary value problem for the exterior of a concave wedge of an arbitrary angle, $W_e=\{0<\Gr<\infty,
0<\Gt<\Ga\}$ ($0<\Ga<\pi$), with the general
impedance boundary conditions 
\beq
\pm \CE_z=\Gn_{\Gr\Gr}^\pm Z\CH_\Gr+\Gn_{\Gr z}^\pm Z \CH_z, \quad 
\mp \CE_\Gr=\Gn_{z\Gr}^\pm Z\CH_\Gr+\Gn_{z z}^\pm Z \CH_z, \quad (\Gr,\Gt)\in S^\pm.
\label{1.2}
\eeq
Next, by applying the Laplace transform {\bf(\ref{dan2})} we reduce the problem
to two sets of six functional equations.
We focus on the case $\Ga=\pi/2$, and in section 3, we transform the  functional equations into two homogeneous vector RHPs
with a $2\times 2$ matrix coefficient.  The residues of the unknown vectors at 
 the geometric optics poles are not prescribed at this stage.
The solution to the second RHP is derived from the solution to the RHP 1 by a 
certain transformation of the impedance parameters and the angle of incidence $\Gb\in(0,\pi)$.
The solution of the RHPs 1 and 2, vectors $\GF^\pm(\Gn)$ and $\hat\GF^\pm(\Gn)$, respectively,
have to
satisfy the symmetry conditions $\GF^+(\Gn)=\GF^-(-\Gn)$ and  $\hat\GF^+(\Gn)=\hat\GF^-(-\Gn)$. 
It is shown that the matrix coefficient of both RHPs
has the structure $G(\Gn)=b_1(\Gn)Q_1(\Gn)+b_2(\Gn)Q_2(\Gn)$, where
$b_1(\Gn)$ and $b_2(\Gn)$ are H\"older functions on the contour of the problem $L=(-\infty,+\infty)$,
and $Q_1(\Gn)$ and $Q_2(\Gn)$ are polynomial $2\times 2$ matrices. This case is algebraic, and the symmetric vector RHPs
can be solved exactly  {\bf(\ref{ant4})}, {\bf(\ref{ant5})}.

In section 4, for simplicity, we assume that the impedance parameters
$\Gn^\pm_{\Gr z}$ and $\Gn^\pm_{z\Gr}$ vanish and deal with 
the boundary conditions (\ref{1.1}).
For this case  we reduce the problem of matrix factorization to a scalar RHP on a genus-3 
Riemann surface.
The coefficient of the RHP on the first and second sheets
coincides with the first and second eigenvalues of the matrix coefficient, respectively.
Because of the symmetry of the problem we manage to solve the associated
Jacobi inversion problem in terms of elliptic integrals and construct
the solution to the matrix factorization
problem by quadratures.

In section 5, we derive  a closed-form solution to the 
vector RHPs.  It turns out that 
the number of free constants in the general solution is governed by  
the location of the zeros $\Gn_j^-$ (RHP 1) and $\Gn_j^+$ (RHP 2) 
of the polynomials 
\beq
\Gd_0(\Gn)=(\Gn^2-k_0^2)\cos^2\Gb-(\Gn-\Gg_1^-)(\Gn-\Gg_4^-), \quad
\hat\Gd_0(\Gn)=(\Gn^2-k_0^2)\cos^2\Gb-(\Gn-\Gg_1^+)(\Gn-\Gg_4^+),
\label{1.3}
\eeq
where $\Gg_1^\pm=k_0(\Gn_{\Gr\Gr}^\pm)^{-1}\sin\Gb$,   
$\Gg_4^\pm=k_0\Gn_{zz}^\pm\sin\Gb$, $k_0=k\sin\Gb$, 
and $k$ is the wave number. 
It is shown that the two RHPs have to fulfill certain compatibility conditions. 
These conditions when satisfied give the solution with two free constants 
regardless  of the location of the zeros $\Gn_j^\pm$.

In section 6, we derive the spectra $S_1(s)$ and $S_2(s)$ of the problem and express the Sommerfeld
integrals through the solution to the two vector RHPs. On continuing analytically
the functions $S_1(s)$ and $S_2(s)$
to the right and to the left, applying the steepest descent method
and the Cauchy theorem,
we recover  the incident, reflected, and diffracted waves and fix the residues of the solution at the geometric optics poles.
For the normal incidence case we find a simple closed-form solution to the RHPs 1 and 2  and verify the compatibility conditions in section 7.

\setcounter{equation}{0}

\section{Formulation: diffraction by a concave wedge of an arbitrary angle}

Consider diffraction of an electromagnetic wave
\beq
\left(\begin{array}{cc}
\CE_z^{inc}, & \CH_z^{inc} \\
\end{array}\right)=
\left(\begin{array}{cc}
E_z^i, & H_z^i \\
\end{array}\right)e^{ikz\cos\Gb-i\Go t},
\label{2.1}
\eeq
where
\beq
\left(\begin{array}{cc}
E_z^{i}, & Z H_z^{i} \\
\end{array}\right)=
\left(\begin{array}{cc}
i_1, & i_2 \\
\end{array}\right)e^{-ik\Gr\cos(\Gt-\Gt_0)\sin\Gb},
\label{2.1'}
\eeq
by a wedge $W=\{0<\Gr<\infty, 2\pi-\Ga<\Gt<2\pi, |z|<\infty\}$, $\Ga\in(0,\pi)$,
characterized by the impedance boundary conditions
$$
\pm \CE_z=\Gn_{\Gr\Gr}^\pm Z\CH_\Gr+\Gn_{\Gr z}^\pm Z \CH_z \quad {\rm on}  \; S^\pm,
$$
\beq
\mp \CE_\Gr=\Gn_{z\Gr}^\pm Z\CH_\Gr+\Gn_{z z}^\pm Z \CH_z\quad {\rm on}  \; S^\pm.
\label{2.2}
\eeq
Here, $k$ ($\I k>0$) is the wave number, $\Go$ is the angular velocity,  $Z=\sqrt{\mu_0/\Gve_0}$ is
 the intrinsic 
impedance of the medium,
$\mu_0$ is the magnetic permeability, $\Gve_0$ is
 the electric permittivity, 
  $\Gn^\pm_{\Gr\Gr}$, $\Gn^\pm_{\Gr z}$, $\Gn^\pm_{z\Gr}$ and $\Gn^\pm_{zz}$ are the impedance parameters.
The symbols $S^\pm$ stand for  the boundaries of the wedge, $S^-=\{ \Gt=0, \Gr>0\}$, and $S^+=\{ \Gt=\Ga, \Gr>0\}$.
Because of the representation (\ref{2.1})
for the incident waves  it is natural to split the electric and 
magnetic fields as
\beq
\left(\begin{array}{cc}
\CE, & \CH \\
\end{array}\right)=
\left(\begin{array}{cc}
E, & H \\
\end{array}\right)e^{ikz\cos\Gb-i\Go t}.
\label{2.3}
\eeq
By employing the Maxwell equations we eliminate the radial components  
of the electric and magnetic fields
$$
E_\Gr=\fr{i}{k\sin^2\Gb}\left[\cos\Gb\fr{\Md}{\Md\Gr}E_z+\fr{Z}{\Gr}\fr{\Md}{\Md\Gt}H_z\right],
$$
\beq
ZH_\Gr=\fr{i}{k\sin^2\Gb}\left[Z\cos\Gb\fr{\Md}{\Md\Gr}H_z-\fr{1}{\Gr}\fr{\Md}{\Md\Gt}E_z\right].
\label{2.4}
\eeq
The resulting boundary conditions, after rearrangement, are formulated in terms of the $z$-components,
$\Gf_1=E_z$ and $\Gf_2=ZH_z$, as
$$
-\fr{i}{\Gr}\fr{\Md\Gf_1}{\Md\Gt} +i\cos\Gb\fr{\Md\Gf_2}{\Md\Gr}\mp\Gg_1^\pm\Gf_1+\Gg_2^\pm\Gf_2=0 \quad {\rm on}  \; S^\pm,
$$
\beq
i\cos\Gb\fr{\Md\Gf_1}{\Md\Gr}+\fr{i}{\Gr}\fr{\Md\Gf_2}{\Md\Gt} +\Gg_3^\pm\Gf_1\pm\Gg_4^\pm\Gf_2=0 \quad {\rm on}  \; S^\pm,
\label{2.5}
\eeq
where
$$
\Gg_1^\pm=\fr{k\sin^2\Gb}{\Gn_{\Gr\Gr}^\pm}, \quad \Gg_2^\pm=\fr{\Gn_{\Gr z}^\pm k\sin^2\Gb}{\Gn_{\Gr\Gr}^\pm}, 
$$
\beq
\Gg_3^\pm=\fr{\Gn_{z\Gr}^\pm k\sin^2\Gb}{\Gn_{\Gr\Gr}^\pm}, \quad \Gg_4^\pm=
\fr{k\sin^2\Gb}{\Gn_{\Gr\Gr}^\pm}(\Gn_{\Gr\Gr}^\pm\Gn_{zz}^\pm-\Gn_{\Gr z}^\pm \Gn_{z\Gr}^\pm).
\label{2.6}
\eeq
It is assumed that the functions $E_z$ and $H_z$ satisfy the  Meixner edge condition
and therefore 
\beq
\Gf_j\to c_j, \quad r\to 0,\quad c_j=\const,\quad j=1,2.
\label{2.6.1}
\eeq
Also, at infinity, they meet the radiation condition
\beq
\Gf_1-E_z^i-E_z^{rw}=O(e^{-\Gve k_0 r}),\quad Z^{-1}\Gf_2-H_z^i-H_z^{rw}=O(e^{-\Gve k_0 r}), \quad 0<\Gt<\Ga,\quad
k_0r\to\infty,
\label{2.6.2}
\eeq
where $\Gve>0$, $k_0=k\sin\Gb$,  $E_z^{rw}$ and $H_z^{rw}$ are the reflected waves.

In the exterior of the wedge, $W_e=\{0<\Gr<\infty, 0<\Gt<\Ga\}$, the functions $\Gf_1$ and $\Gf_2$ 
satisfy the Helmholtz equation. To convert the boundary value problem for the Helmholtz equation to a system of functional
equations, we use the scheme {\bf(\ref{dan2})}.
On making the affine transformation of the Cartesian coordinates
$u=x-y\cot\Ga$ and $v=y\csc\Ga$ we obtain that, in the new coordinates, the functions
$\Gf_1$ and $\Gf_2$ are the solutions of the following differential equation in a quarter-plane:
\beq
\left(\fr{\Md^2}{\Md u^2}+\fr{\Md^2}{\Md v^2}-2\cos\Ga\fr{\Md^2}{\Md u\Md v}+k_0^2\sin^2\Ga\right) \Gf_j 
=0, \quad 0<u,v<\infty,\quad j=1,2,
\label{2.7}
\eeq
subject to the four boundary conditions
$$
 i\left(\csc\Ga\fr{\Md}{\Md u}-\cot\Ga\fr{\Md}{\Md v}\right)\Gf_1+i\cos\Gb\fr{\Md\Gf_2}{\Md v}-\Gg_1^+\Gf_1+\Gg_2^+\Gf_2=0,
\quad (u,v)\in\tilde S^+,
$$
$$
i\cos\Gb\fr{\Md\Gf_1}{\Md v}- i\left(\csc\Ga\fr{\Md}{\Md u}-\cot\Ga\fr{\Md}{\Md v}\right)\Gf_2
 +\Gg_3^+\Gf_1+\Gg_4^+\Gf_2=0, \quad (u,v)\in\tilde S^+,
$$
$$
 i\left(-\csc\Ga\fr{\Md}{\Md v}+\cot\Ga\fr{\Md}{\Md u}\right)\Gf_1+i\cos\Gb\fr{\Md\Gf_2}{\Md u}+\Gg_1^-\Gf_1+\Gg_2^-\Gf_2=0,
\quad (u,v)\in\tilde S^-,
$$
\beq
i\cos\Gb\fr{\Md\Gf_1}{\Md u}+ i\left(\csc\Ga\fr{\Md}{\Md v}-\cot\Ga\fr{\Md}{\Md u}\right)\Gf_2
 +\Gg_3^-\Gf_1-\Gg_4^-\Gf_2=0, \quad (u,v)\in\tilde S^-,
\label{2.8}
\eeq
where  $\tilde S^+=\{(u,v)\in {\Bbb R}^2 | u=0, v>0\}$, $\tilde S^-=\{(u,v)\in {\Bbb R}^2 | v=0, u>0\}$.

By means of the Laplace transform
\beq  
\tilde\Gf_j(u,\Gn)=\int\limits_0^\infty e^{i\Gn v}\Gf_j(u,v)dv
\label{2.9}
\eeq
the differential equation (\ref{2.7}) may be put into the form
\beq
\left(\fr{d^2}{du^2}+2i\Gn\cos\Ga\fr{d}{du}+k_0^2\sin^2\Ga-\Gn^2\right)\tilde\Gf_j(u,\Gn)=f_j(u), \quad j=1,2,
\label{2.10}
\eeq
where
\beq
f_j(u)=\fr{\Md\Gf_j}{\Md v}(u,0)-i\Gn\Gf_j(u,0)-2\cos\Ga\fr{\Md\Gf_j}{\Md u}(u,0).
\label{2.11}
\eeq
The two roots of the characteristic equation of the differential operator in (\ref{2.10}) are $-i\Gn\cos\Ga\pm\sqrt{\Gn^2-k_0^2}\sin\Ga$.
Fix a branch $\Gz=\sqrt{\Gn^2-k_0^2}$ of the two-valued function $\Gz^2=\Gn^2-k_0^2$ by the 
condition $\Gz(0)=-ik_0$. The branch is a single-valued analytic function in the $\Gn$-plane cut along the
line joining the branch points $\pm k_0$ and passing through the infinite point.
As $-\infty<\Gn<+\infty$, the branch chosen possesses the property $\R\Gz\ge 0$, and the general solution to equation (\ref{2.10}) bounded as $u\to\infty$
 has the form
\beq
\tilde\Gf_j(u,\Gn)=A_j(\Gn)e^{-q u}-\fr{1}{2q}\int_0^\infty e^{-q|u-u_1|}f_j(u_1)du_1,\quad j=1,2,
\label{2.12}
\eeq
where $q=-i\Gn\cos\Ga+\Gz\sin\Ga$ ($\R q\ge 0)$.
 To derive a functional equation for $\tilde\Gf_j$, we apply the Laplace transform with respect to $u$
using the function $q$ as its parameter
\beq
\hat\Gf_j(iq,v)=\int_0^\infty e^{-q u}\Gf_j(u,v)du.
\label{2.13}
\eeq
On referring to (\ref{2.11}), we find that
\beq
\int_0^\infty f_j(u)e^{-q u}du=\fr{d}{dv}\hat\Gf_j(iq,0)-(i\Gn+2q\cos\Ga)\hat\Gf_j(iq,0)+2\cos\Ga\Gf_j^\circ,
\label{2.14}
\eeq
where $\Gf_j^\circ=\Gf_j(0,0)$.
On the other hand, from (\ref{2.12})
$$
\tilde\Gf_j(0,\Gn)= A_j(\Gn)-\fr{1}{2q}\int_0^\infty e^{-q u}f_j(u)du,
$$
\beq
\fr{d\tilde\Gf_j}{du}(0,\Gn)=-q A_j(\Gn)-\fr12\int_0^\infty e^{-q u}f_j(u)du.
\label{2.15}
\eeq
By excluding $A_j(\Gn)$ and the Laplace transforms of the functions $f_j(u)$ and employing
the second relation in (\ref{2.15}) we get the following two equations:
\beq
\fr{d\tilde\Gf_j}{du}(0,\Gn)+q\tilde\Gf_j(0,\Gn)+\fr{d\hat\Gf_j}{dv}(iq,0)-(i\Gn+2q\cos\Ga)\hat\Gf_j(iq,0)+2\cos\Ga\Gf_j^\circ=0\quad j=1,2.
\label{2.16}
\eeq
These two equations should be complemented by the Laplace-transformed boundary conditions. We apply the Laplace transform
(\ref{2.9}) to the boundary conditions (\ref{2.8}) on the boundary $\tilde S^+$ and the transform (\ref{2.13})
to the conditions on the side $\tilde S^-$. We have
$$
i\csc\Ga\fr{d\tilde\Gf_1}{du}(0,\Gn)-(\Gg_1^++\Gn\cot\Ga)\tilde\Gf_1(0,\Gn)+(\Gg_2^++\Gn\cos\Gb)\tilde\Gf_2(0,\Gn)
+i\cot\Ga\Gf_1^\circ-i\cos\Gb\Gf_2^\circ=0,
$$$$
-i\csc\Ga\fr{d\tilde\Gf_2}{du}(0,\Gn)+(\Gg_3^++\Gn\cos\Gb)\tilde\Gf_1(0,\Gn)
+(\Gg_4^++\Gn\cot\Ga)\tilde\Gf_2(0,\Gn)-i\cos\Gb\Gf_1^\circ-i\cot\Ga\Gf_2^\circ=0,
$$$$
-i\csc\Ga\fr{d\hat\Gf_1}{dv}(iq,0)+(\Gg_1^-+iq\cot\Ga)\hat\Gf_1(iq,0)
+(\Gg_2^-+iq\cos\Gb)\hat\Gf_2(iq,0)-i\cot\Ga\Gf_1^\circ-i\cos\Gb\Gf_2^\circ=0,
$$
\beq
i\csc\Ga\fr{d\hat\Gf_2}{dv}(iq,0)+(\Gg_3^-+iq\cos\Gb)\hat\Gf_1(iq,0)
-(\Gg_4^-+iq\cot\Ga)\hat\Gf_2(iq,0)-i\cos\Gb\Gf_1^\circ+i\cot\Ga\Gf_2^\circ=0.
\label{2.17}
\eeq
Notice that by applying the Laplace transform (\ref{2.9}) with respect to 
$u$ and then the Laplace transform (\ref{2.13}) with respect to $v$
we obtain another set of six equations which coincide with (\ref{2.16}) and (\ref{2.17})
if we make the following transformation of the parameters and 
the functions:
$$
\Gg_1^+\leftrightarrow\Gg_1^-,\quad \Gg_2^+\leftrightarrow-\Gg_2^-,\quad
\Gg_3^+\leftrightarrow-\Gg_3^-,\quad \Gg_4^+\leftrightarrow\Gg_4^-,\quad
\Gb\leftrightarrow\pi-\Gb,
$$$$
\tilde\Gf_j(0,\Gn)\leftrightarrow\hat\Gf_j(\Gn,0),\quad
\fr{d\tilde\Gf_j}{du}(0,\Gn)\leftrightarrow\fr{d\hat\Gf_j}{dv}(\Gn,0),
$$
\beq
\hat\Gf_j(iq,0)\leftrightarrow\tilde\Gf_j(0,iq),\quad
\fr{d\hat\Gf_j}{dv}(iq,0)\leftrightarrow\fr{d\tilde\Gf_j}{du}(0,iq),\quad j=1,2.
\label{2.18}
\eeq
In the next section we consider the case $\Ga=\pi/2$ and
 reduce the system of six functional equations (\ref{2.16}), (\ref{2.17}) and the one obtained from the system  (\ref{2.16}), (\ref{2.17}) by the transformation (\ref{2.18})
to two symmetric RHPs with a $2\times 2$ matrix coefficient.

\section{Vector RHPs in the case of a right-angled concave wedge}

\begin{figure}[t]
\centerline{
\scalebox{0.6}{\includegraphics{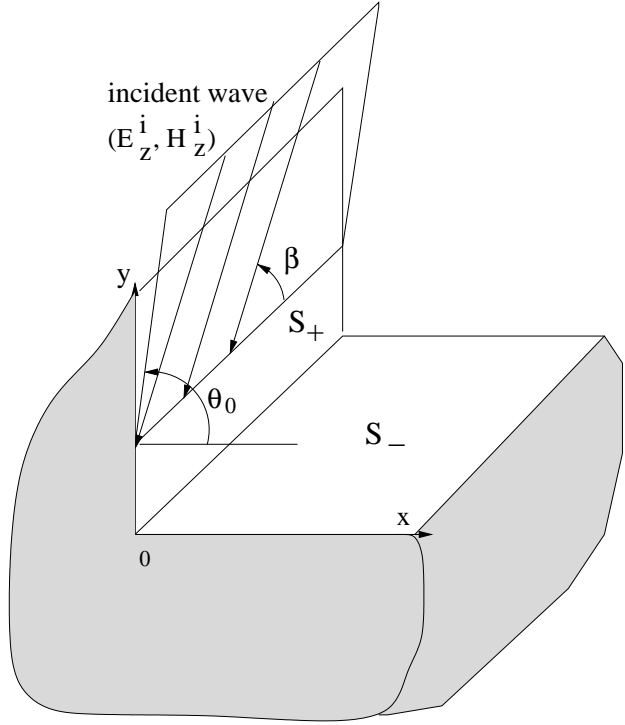}}}
\caption{Geometry of the problem on a right-angled concave wedge.}
\label{fig1}
\end{figure} 
\setcounter{equation}{0}

\subsection{Statement of the problem}

 If the domain of interest is a quarter-plane  (Fig. 1),  then the standard Cartesian 
coordinates $x=u$, $y=v$ can be employed, and $q=\Gz$, $\Gz=\sqrt{\Gn^2-k_0^2}$.
From the boundary conditions (\ref{2.17}) we express the derivatives $d\tilde\Gf_j(0,\Gn)/dx$ and 
$d\hat\Gf_j(i\Gz,0)/dy$
through the functions $\tilde\Gf_j(0,\Gn)$ and $\hat\Gf_j(i\Gz,0)$ ($j=1,2$)
$$
\fr{d\hat\Gf_1}{dy}(i\Gz,0)=-i\Gg^-_1\hat\Gf_1(i\Gz,0)-(i\Gg_2^--\Gz\cos\Gb)\hat\Gf_2(i\Gz,0)-\Gf_2^\circ\cos\Gb,
$$
$$
\fr{d\hat\Gf_2}{dy}(i\Gz,0)=(i\Gg_3^--\Gz\cos\Gb)\hat\Gf_1(i\Gz,0)-i\Gg^-_4\hat\Gf_2(i\Gz,0)+\Gf_1^\circ\cos\Gb,
$$
$$
\fr{d\tilde\Gf_1}{dx}(0,\Gn)=-i\Gg_1^+\tilde\Gf_1(0,\Gn)+i(\Gg_2^++\Gn\cos\Gb)\tilde\Gf_2(0,\Gn)+\Gf_2^\circ\cos\Gb,
$$
\beq
\fr{d\tilde\Gf_2}{dx}(0,\Gn)=-i(\Gg_3^++\Gn\cos\Gb)\tilde\Gf_1(0,\Gn)-i\Gg_4^+\tilde\Gf_2(0,\Gn)-\Gf_1^\circ\cos\Gb.
\label{3.0.3}
\eeq
Next,  on replacing $\Gn$ 
by $-\Gn$ in (\ref{2.16}) we obtain 
two more equations
\beq
\fr{d\tilde\Gf_j}{dx}(0,-\Gn)+\Gz\tilde\Gf_j(0,-\Gn)+\fr{d\hat\Gf_j}{dy}(i\Gz,0)+i\Gn\hat\Gf_j(i\Gz,0)=0,\quad j=1,2.
\label{3.1}
\eeq
These two extra equations in conjunction with (\ref{2.16}) allow us to express the functions $\hat\Gf_j(i\Gz,0)$
through the functions $\tilde\Gf_j(0,\pm\Gn)$
$$
\hat\Gf_1(i\Gz,0)=-\fr{\Gg_1^++i\Gz}{2\Gn}[\tilde\Gf_1(0,\Gn)-\tilde\Gf_1(0,-\Gn)]+
\fr{\Gg_2^++\Gn\cos\Gb}{2\Gn}\tilde\Gf_2(0,\Gn)+\fr{-\Gg_2^++\Gn\cos\Gb}{2\Gn}\tilde\Gf_2(0,-\Gn),
$$
\beq
\hat\Gf_2(i\Gz,0)=-\fr{\Gg_3^++\Gn\cos\Gb}{2\Gn}\tilde\Gf_1(0,\Gn)+\fr{\Gg_3^+-\Gn\cos\Gb}{2\Gn}\tilde\Gf_1(0,-\Gn)
-\fr{\Gg_4^++i\Gz}{2\Gn}[\tilde\Gf_2(0,\Gn)-\tilde\Gf_2(0,-\Gn)].
\label{3.2}
\eeq
On substituting the expressions for the functions $\hat\Gf_j(i\Gz,0)$ and the derivatives $d\tilde\Gf_j(0,\Gn)/dx$ and 
$d\hat\Gf_j(i\Gz,0)/dy$ 
 in equations  (\ref{2.16}) we ultimately obtain
\beq
A(\Gn)\GF^+(\Gn)+B(\Gn)\GF^+(-\Gn)=0, \quad \Gn\in L=(-\infty,+\infty),
\label{3.3}
\eeq
where
$$
\GF^+(\Gn)=
\left(\begin{array}{c}
\tilde\Gf_1(0,\Gn)\\
\tilde\Gf_2(0,\Gn)\\
\end{array}\right), \quad A(\Gn)=\left(\begin{array}{cc}
a_{11}(\Gn)   & a_{12}(\Gn) \\
a_{21}(\Gn)   & a_{22}(\Gn) \\
\end{array}\right), \quad B(\Gn)=\left(\begin{array}{cc}
b_{11}(\Gn)   & b_{12}(\Gn) \\
b_{21}(\Gn)   & b_{22}(\Gn) \\
\end{array}\right), 
$$
$$
a_{11}=\Gd_{1+}(\Gn)[1-\Gd_{1-}(\Gn)]-\Gd_{2-}(\Gn)\Gd_{3+}(\Gn), \quad 
a_{12}=-\Gd_{2+}(\Gn)[1-\Gd_{1-}(\Gn)]-\Gd_{2-}(\Gn)\Gd_{4+}(\Gn), 
$$$$
a_{21}=\Gd_{3+}(\Gn)[1-\Gd_{4-}(\Gn)]+\Gd_{3-}(\Gn)\Gd_{1+}(\Gn), \quad 
a_{22}=\Gd_{4+}(\Gn)[1-\Gd_{4-}(\Gn)]-\Gd_{3-}(\Gn)\Gd_{2+}(\Gn), 
$$$$
b_{11}=\Gd_{1-}(\Gn)\Gd_{1+}(-\Gn)+\Gd_{2-}(\Gn)\Gd_{3+}(-\Gn), \quad 
b_{12}=-\Gd_{1-}(\Gn)\Gd_{2+}(-\Gn)+\Gd_{2-}(\Gn)\Gd_{4+}(-\Gn), 
$$$$
b_{21}=-\Gd_{3-}(\Gn)\Gd_{1+}(-\Gn)+\Gd_{4-}(\Gn)\Gd_{3+}(-\Gn), \quad 
b_{22}=\Gd_{3-}(\Gn)\Gd_{2+}(-\Gn)+\Gd_{4-}(\Gn)\Gd_{4+}(-\Gn),  
$$
$$
\Gd_{j+}(\Gn)=\Gg_j^++i\Gz, \quad \Gd_{j-}(\Gn)=\fr{\Gg_j^-+\Gn}{2\Gn}, \quad j=1,4,
$$
\beq
\Gd_{j+}(\Gn)=\Gg_j^++\Gn\cos\Gb, \quad \Gd_{j-}(\Gn)=\fr{\Gg_j^-+i\Gz\cos\Gb}{2\Gn}, \quad j=2,3.
\label{3.4}
\eeq
It can be directly verified that $A(\Gn)=B(-\Gn)$ for all $\Gn$, and the replacement  of $\Gn$ by $-\Gn$ does not change the problem (\ref{3.3}).
We now summarize the results.  Denote  $\Gn_0=k_0\sin\Gt_0$,  $\hat\Gn_0=k_0\cos\Gt_0$.

\vspace{.1in}

{\sl Theorem 3.1. Suppose that $\Ga=\pi/2$, and the incident wave is $(E_z^{i}, H_z^{i})$. Then the diffraction problem (\ref{2.7}), (\ref{2.8}), (\ref{2.6.1}), (\ref{2.6.2}) is equivalent to
the following symmetric vector RHP provided its solution  recovers the incident and reflected waves.

RHP 1. Find two vectors  $\GF^\pm(\Gn)$ analytic everywhere in the half-planes ${\Bbb C}^\pm=\{\pm\I\Gn>0\}$ 
except at the simple poles $\Gn=\pm\Gn_0$
and H\"older-continuous up the real axis $L$. At the contour $L$, their limit values satisfy the boundary condition 
\beq
\GF^+(\Gn)=G(\Gn)\GF^-(\Gn), \quad \Gn\in L,
\label{3.6}
\eeq
where $G(\Gn)=-[A(\Gn)]^{-1}B(\Gn)$. In the plane, the vectors meet the symmetry condition 
\beq
\GF^-(\Gn)=\GF^+(-\Gn), \quad \Gn\in{\Bbb C}^-,
\label{3.4'}
\eeq
and at infinity, they vanish.
}

\vspace{.1in}

On examining
the coefficient of the vector RHP we discover that it is a
 nonsingular matrix admitting the representation
\beq
G(\Gn)=\fr{1}{d(\Gn)}[G^0(\Gn)+\Gz G^1(\Gn)],
\label{3.7}
\eeq
where $d(\Gn)$ is a scalar, $G^0$ and  $G^1$ are  polynomial matrices.
 Any nonsingular matrix $G$ of such a structure
can be factorized  {\bf(\ref{ant4})} by reducing the vector RHP to a scalar RHP
on a Riemann surface of the associated algebraic function. 
To simplify our derivations, it will be helpful to assume that either $\Gg_2^\pm=\Gg_3^\pm=0$ and the angle $\Gb$ is arbitrary ($0<\Gb<\pi$),
or  that all the four impedance parameters $\Gg_j^\pm$ are arbitrary and $\Gb=\pi/2$.  In what follows we consider the former case, namely, $\Gg_2^\pm=\Gg_3^\pm=0$ and
$\Gb\in(0,\pi)$.

We also notice that the transformation (\ref{2.18}) maps the diffraction problem to another vector RHP.

{\it RHP 2. Find two vectors  $\hat\GF^\pm(\Gn)=(\hat\Gf_1(\pm\Gn,0),\hat\Gf_2(\pm\Gn,0))^T$ 
analytic everywhere in the half-planes ${\Bbb C}^\pm$ 
except at the simple poles $\Gn=\pm\hat\Gn_0$
and H\"older-continuous up the real axis $L$. At the contour $L$, their limit values satisfy the boundary condition 
\beq
\hat\GF^+(\Gn)=\hat G(\Gn)\hat\GF^-(\Gn), \quad \Gn\in L,
\label{3.6'}
\eeq
where $\hat G(\Gn)$
is the matrix $G(\Gn)$ whose parameters $\Gg_j^\pm$ ($j=1,\ldots,4$) and $\Gb$ are transformed 
by (\ref{2.18}). 
At infinity, the vectors $\hat\GF^+(\Gn)$ and $\hat\GF^-(\Gn)$ vanish, and in the plane, they  meet the symmetry condition $\hat\GF^-(\Gn)=\hat\GF^+(-\Gn),  \Gn\in{\Bbb C^-}$.}

Undoubtedly, by the symmetry transformation (\ref{2.18}) any result obtained for the first vector RHP  is valid for the second problem and vice versa. Since both problems 
recover the same vectors, the values of the Laplace-transformed functions $\Gf_1$ and $\Gf_2$ on the faces of the wedge, we solve 
only one problem,  the vector RHP 1 stated in Theorem 3.1 (we will call it the vector RHP).

\subsection{Reflected waves}

Before we start the matrix factorization procedure we note that the choice 
of the residues of the functions $\GF^\pm(\Gn)$ and $\hat\GF^\pm(\Gn)$ at the geometric optics poles $\Gn=\pm\Gn_0$ and $\pm\hat\Gn_0$, respectively, have to bring us to
 the incident and reflected waves. To recover them, 
split the incident waves $E_z^i$ and $H_z^i$ into two waves
$$
\left(\begin{array}{cc}
E_z^{i+}, & Z H_z^{i+} \\
\end{array}\right)=
\left(\begin{array}{cc}
i_1, & i_2 \\
\end{array}\right)e^{-ik_0\Gr\cos(\Gt-\Gt_0)}\Go(\Gt;\Gt_0,\pi/2),
$$
\beq
\left(\begin{array}{cc}
E_z^{i-}, & Z H_z^{i-} \\
\end{array}\right)=
\left(\begin{array}{cc}
i_1, & i_2 \\
\end{array}\right)e^{-ik_0\Gr\cos(\Gt-\Gt_0)}\Go(\Gt;0,\Gt_0),
\label{3.0.1}
\eeq
where
\beq
\Go(\Gt;a,b)=\left\{\begin{array}{cc}
1, & a<\Gt<b,\\
0 & {\rm otherwise.}\\
\end{array}
\right.
\label{3.0.2}
\eeq
The first wave $(E_z^{i+}, H_z^{i+})$ reflects from the vertical wall $\{x=0, 0<y<\infty\}$, falls on the horizontal wall
$\{0<x<\infty, y=0\}$, reflects from it and runs away to infinity. 
We denote the first and second reflected waves as $(E_{z+}^r, H_{z+}^r)$ and $(E_{z+}^{R}, H_{z+}^{R})$, respectively, 
which are
$$
\left(\begin{array}{cc}
E_{z+}^{r}, & Z H_{z+}^{r} \\
\end{array}\right)=
\left(\begin{array}{cc}
r_1^+, & r_2^+ \\
\end{array}\right)e^{i\hat\Gn_0 x-i\Gn_0y},
$$
\beq
\left(\begin{array}{cc}
E_{z+}^{R}, & ZH_{z+}^{R} \\
\end{array}\right)=
\left(\begin{array}{cc}
R_1^+, & R_2^+ \\
\end{array}\right)e^{i\hat\Gn_0 x+i\Gn_0y}\Go(\Gt;0,\Gt_0).
\label{3.1.1}
\eeq
Likewise, the second wave $(E_z^{i-}, H_z^{i-})$ impinges upon the horizontal face of the wedge, reflects  from it, strikes
 the vertical wall and then goes to infinity. The
reflection waves are described by 
$$
\left(\begin{array}{cc}
E_{z-}^{r}, & ZH_{z-}^{r} \\
\end{array}\right)=
\left(\begin{array}{cc}
r_1^-, & r_2^- \\
\end{array}\right)e^{-i\hat\Gn_0 x+i\Gn_0y},
$$
\beq
\left(\begin{array}{cc}
E_{z-}^{R}, & ZH_{z-}^{R} \\
\end{array}\right)=
\left(\begin{array}{cc}
R_1^-, & R_2^- \\
\end{array}\right)e^{i\hat\Gn_0 x+i\Gn_0y}\Go(\Gt;\Gt_0,\pi/2).
\label{3.1.2}
 \eeq
The reflection coefficients  are recovered by substituting the incident and reflected waves (\ref{3.0.1}),  (\ref{3.1.1}) and  (\ref{3.1.2}) into (\ref{2.8})
$$
r_1^+=K_{1}^-i_1+K_{2}i_2, \quad r_2^+=-K_{2}i_1+K_{1}^+i_2, 
$$$$
  R_1^+=\hat K_{1}^-r_1^++\hat K_{2}r_2^+, \quad R_2^+=-\hat K_{2}r_1^++\hat K_{1}^+r_2^+, 
$$
$$
r_1^-=\hat K_{1}^-i_1-\hat K_{2}i_2, \quad r_2^-=\hat K_{2}i_1+\hat K_{1}^+i_2, 
$$
\beq
R_1^-=K_{1}^-r_1^--K_{2}r_2^-, \quad R_2^-=K_{2}r_1^-+K_{1}^+r_2^-, 
\label{3.7.1}
\eeq
where
$$
K_{1}^\pm=\fr{(\hat\Gn_0\pm\Gg_1^+)(\hat\Gn_0\mp\Gg_4^+)-\Gn_0^2\cos^2\Gb}{\GD_0}, \quad K_{2}=\fr{2\Gn_0\hat\Gn_0\cos\Gb}{\GD_0},
$$$$
\hat K_{1}^\pm=\fr{(\Gn_0\pm\Gg_1^-)(\Gn_0\mp\Gg_4^-)-\hat\Gn_0^2\cos^2\Gb}{\hat\GD_0}, \quad \hat K_{2}=\fr{2\Gn_0\hat\Gn_0\cos\Gb}{\hat\GD_0},
$$
\beq
\GD_0=(\hat\Gn_0+\Gg_1^+)(\hat\Gn_0+\Gg_4^+)+\Gn_0^2\cos^2\Gb,\quad \hat\GD_0=(\Gn_0+\Gg_1^-)(\Gn_0+\Gg_4^-)+\hat\Gn_0^2\cos^2\Gb.
\label{3.7.2}
\eeq

\subsection{Analysis of the matrix $G(\Gn)$}

The matrix
$G(\Gn)$ is continuous and nonsingular everywhere in the contour $L$, and as $\Gn\to\pm\infty$,
$G(\Gn)\sim \diag\{-1,-1\}$. Compute first the determinant of $G$
\beq
\det G(\Gn)=\fr{\Gd_0(-\Gn)}{\Gd_0(\Gn)},
\label{3.7'''}
\eeq
where
\beq
\Gd_0(\Gn)=(\Gn^2-k_0^2)\cos^2\Gb-(\Gn-\Gg_1^-)(\Gn-\Gg_4^-),
\label{3.7''''}
\eeq
and determine its index 
\beq
2\Gk=\fr{1}{2\pi}[\arg \det G(\Gn)]_{-\infty}^\infty.
\label{3.7''}
\eeq
We distinguish the following three cases:

(i) the two zeros of the quadratic polynomial $\Gd_0(\Gn)$ lie in the lower half-plane, $\Gd_0(\Gn)=-(\Gn+\tau_1)(\Gn+\tau_2)\sin^2\Gb$, $\I\tau_j>0$, $j=1,2$,

(ii) the zeros lie in the opposite half-planes,
$\Gd_0(\Gn)=-(\Gn-\tau_1)(\Gn+\tau_2)\sin^2\Gb$, $\I\tau_j>0$, $j=1,2$, and

(iii) both zeros lie in the upper half-plane, $\Gd_0(\Gn)=-(\Gn-\tau_1)(\Gn-\tau_2)\sin^2\Gb$, $\I\tau_j>0$, $j=1,2$.

By the argument principle, $\Gk$ is an integer number, 
and in the case (i) $\Gk=1$ (in the upper half-plane,
the function $\Gd_0(-\Gn)$ has two zeros, while the function $\Gd_0(\Gn)$ has no zeros in 
${\Bbb C^+}$). In the case (ii)
$\Gk=0$, and in the last case $\Gk=-1$.

Our next step is to rearrange the representation (\ref{3.7}) as
\beq
G(\Gn)=\fr{\Gz}{\Gd_0(\Gn)\Gd_1(\Gn)}G^1(\Gn)\left[I+\fr{1}{\Gz d_1(\Gn)}\left(\begin{array}{cc}
r_1(\Gn) \; & r_2(\Gn) \\
r_3(\Gn) \; & r_4(\Gn) \\
\end{array}
\right)\right],
\label{3.8}
\eeq
where $I$ is the unit $2\times 2$ matrix, $r_j(\Gn)$ ($j=1,\ldots,4$) are polynomials, 
\beq
\Gd_1(\Gn)=(\Gg_{1}^++i\Gz)(\Gg_{4}^++i\Gz)+\Gn^2\cos^2\Gb,
\label{3.9}
\eeq
the entries $g^1_{st}(\Gn)$ of the polynomial matrix $G^1(\Gn)$
are
$$
g^1_{jj}(\Gn)=-i(\Gg_1^+ + \Gg_4^+)[k_0^2\cos^2\Gb-\Gn^2\sin^2\Gb+(-1)^j\Gn(\Gg_1^--\Gg_4^-)+  \Gg_1^-\Gg_4^-],\quad j=1,2,
$$
$$
g^1_{12}(\Gn)=-2i\Gn\cos\Gb[\Gg_1^- \Gg_4^- -(\Gg_4^+)^2 -k_0^2\sin^2\Gb],
$$
\beq
g^1_{21}(\Gn)=2i\Gn\cos\Gb[\Gg_1^- \Gg_4^- -(\Gg_1^+)^2 -k_0^2\sin^2\Gb],
\label{3.10}
\eeq
and $d_1(\Gn)=\det G^1(\Gn)$ is a degree-4 even polynomial
\beq
d_1(\Gn)=\Ga_0+\Ga_1\Gn^2+\Ga_2\Gn^4
\label{3.14}   
\eeq
whose coefficients are
$$
\Ga_0=-(\Gg_1^+ + \Gg_4^+)^2 ( \Gg_1^-\Gg_4^- + k_0^2\cos^2 \Gb)^2, \quad \Ga_2=-(\Gg_1^++\Gg_4^+)^2\sin^4\Gb,
$$
$$
\Ga_1 = (\Gg_1^++\Gg_4^+)^2 [(\Gg_1^-)^2+(\Gg_4^-)^2-2k_0^2\cos^4\Gb] + 
  \{-4 (\Gg_1^- \Gg_4^-)^2 - (2 \Gg_1^+ \Gg_4^+ - k_0^2)^2
  $$
  \beq
   +  2 \Gg_1^-\Gg_4^- [(\Gg_1^+ -\Gg_4^+)^2 + 2 k_0^2] + 
     2 k_0^2 [(\Gg_1^+)^2 - 2 \Gg_1^- \Gg_4^- + (\Gg_4^+)^2 + k_0^2] \cos 2 \Gb - 
     k_0^4 \cos^2 2\Gb\}\cos^2\Gb.
     \label{3.15}
\eeq
It is an easy matter now to represent the matrix $G(\Gn)$
in the form
\beq
G(\Gn)=\fr{\Gd_*\sqrt{\GD(\Gn)}}{\Gd_0(\Gn)}G^1(\Gn)\GG(\Gn),
\label{3.11}
\eeq
where $\Gd_*$ is a constant to be determined,
$$
\GG(\Gn)=\left(\begin{array}{cc}
b_0(\Gn)+c_0(\Gn)l(\Gn) \; & c_0(\Gn)m(\Gn) \\
c_0(\Gn)n(\Gn) \; & b_0(\Gn)-c_0(\Gn)l(\Gn) \\
\end{array}
\right),
$$$$
b_0=\fr{b}{\sqrt{\GD}},\quad c_0=\fr{c}{\sqrt{\GD}},\quad
\GD=b^2-c^2f,\quad f=l^2+mn,
$$
$$
b=\fr{\Gz}{\Gd_*\Gd_1}\left(1+\fr{r}{\Gz d_1}\right), \quad 
c=\fr{i\hat\Gg\Gn\cos\Gb}{\Gd_*\Gd_1 d_1},
$$
\beq
\hat\Gg=\Gg_1^-\Gg_4^-+\Gg_1^+\Gg_4^+-k_0^2\sin^2\Gb,
\label{3.12}
\eeq
and $l$, $m$, $n$ and $r$ are polynomials,
$$
l(\Gn)=\Gn  \cos\Gb [\Gn^2 (\Gg_1^+ - \Gg_4^+) - 
   2 \Gn (\Gg_1^- - \Gg_4^-) (\Gg_1^+ 
   + \Gg_4^+)
   $$$$ 
   + (\Gg_1^+ - \Gg_4^+) (2 \Gg_1^-\Gg_4^- - 
      3 k_0^2) - (\Gg_1^+ - \Gg_4^+)( \Gn^2-k_0^2) \cos 2 \Gb],
$$
$$
m(\Gn)=
-\fr34 \Gn^4 +  \Gn^3 (\Gg_1^- - \Gg_4^-)
  + 2\Gn  (\Gg_1^- - \Gg_4^-) [(\Gg_4^+)^2 - k_0^2] 
 $$
$$
     +[(\Gg_4^+)^2 - k_0^2](2 \Gg_1^- \Gg_4^- + k_0^2) + 
    \fr34 \Gn^2 [4 \Gg_1^- \Gg_4^- - 4 (\Gg_4^+)^2 + k_0^2]- \fr14\Gn^2( \Gn^2-k_0^2)\cos4\Gb
$$
\beq    
  + \{\Gn^4 - \Gn^3 (\Gg_1^- - \Gg_4^-) + \Gn^2 [\Gg_1^- \Gg_4^- - (\Gg_4^+)^2 - k_0^2] + 
    k_0^2 [(\Gg_4^+)^2 - k_0^2]\} \cos 2 \Gb,
\label{3.13}
\eeq    
the polynomial $n(\Gn)$ coincides with $-m(\Gn)$ if $\Gg_1^-$ and $\Gg_1^+$ are replaced by $\Gg_4^-$ and $\Gg_4^+$, respectively, and
\beq
r(\Gn)=\fr{i}{16}  (\Gg_1^+ + \Gg_4^+) [-\Gn^2 + 2 (\Gg_1^+ \Gg_4^+ + k_0^2) + 
   \Gn^2 \cos 2 \Gb] 
   $$$$
   \times[3 \Gn^4 - 8\Gn^2 (\Gg_1^-)^2 - 8 \Gn^2 \Gg_1^- \Gg_4^ - 
   -8 \Gn^2 (\Gg_4^-)^2 
   + 8 (\Gg_1^-)^2 (\Gg_4^-)^2 + 2 \Gn^2 k_0^2
   $$$$ + 8 \Gg_1^- \Gg_4^- k_0^2 + 3 k_0^4 - 
   4 (\Gn^2 - k_0^2) (\Gn^2 + 2 \Gg_1^- \Gg_4^- + k_0^2) \cos 2\Gb + (\Gn^2 - k_0^2)^2 \cos 4\Gb].
\label{3.13A}
\eeq    
In (\ref{3.12}) we divided the coefficients of the matrix $\GG(\Gn)$ by $\sqrt{\GD(\Gn)}$ ($\sqrt{\GD(\Gn)}$
is a fixed branch of the function $\GD^{1/2}(\Gn)$) and in (\ref{3.11}) 
we multiplied the matrix $\GG(\Gn)$ by the same factor. This transformation implies $\det\GG(\Gn)=1$
for all $\Gn\in L$ and simplifies the procedure of matrix factorization.
Notice that from (\ref{3.11}), 
\beq
\det G(\Gn)=\fr{\Gd_*^2\GD(\Gn)d_1(\Gn)}{\Gd_0^2(\Gn)}.
\label{3.12'}
\eeq
On comparing (\ref{3.12'}) and (\ref{3.7'''}), we obtain that 
$\GD(\Gn)$ is a rational even function
\beq
\GD(\Gn)=\fr{\Gd_0(\Gn)\Gd_0(-\Gn)}{\Gd_*^2d_1(\Gn)}.
\label{3.12''}
\eeq

\setcounter{equation}{0}

\section{Scalar RHP on a Riemann surface}

\subsection{Statement of the problem}

Since $G^1(\Gn)$ is a polynomial matrix, the problem of factorization of the matrix $G(\Gn)$ reduces to
the one for the matrix $\GG(\Gn)$. Its factorization can be expressed through the solution of the 
associated scalar RHP on the genus-3 Riemann surface $\frak R$ of the algebraic function
$w^2=f(\Gn)$.
The function $f(\Gn)$  is an even degree-8 polynomial which has 
the form
\beq
f(\Gn)=h_0+h_1\Gn^2+h_2\Gn^4+h_3\Gn^6+h_4\Gn^8,
\label{3.16}
\eeq
where
$$
h_0=-4u_1 u_2 u_3^2, \quad h_4=-4\sin^8\Gb,
$$$$
h_1=
4(\Gg_1^- - \Gg_4^-)^2u_1 u_2 + 
    (\Gg_1^+ - \Gg_4^+)^2
     (2 \Gg_1^- \Gg_4^- - 3 k_0^2 + k_0^2 \cos 2\Gb)^2  \cos^2\Gb -2 u_3 
 $$$$
\times\{ u_2 [(\Gg_1^- \Gg_4^- - (\Gg_1^+)^2)  (\cos 2\Gb+3) + 2k_0^2 \sin^4\Gb] 
+u_1[(\Gg_1^- \Gg_4^- - (\Gg_4^+)^2) (\cos 2\Gb+3) + 2k_0^2 \sin^4\Gb]\},
 $$$$
h_2=4 u_4^+ u_4^-
+8(\Gg_1^-\Gg_4^-u_4^++\Gg_1^+\Gg_4^+u_4^-)\cos^2\Gb+8\Gg_1^-\Gg_1^+\Gg_4^-\Gg_4^+\cos^2\Gb(\cos 2\Gb-3)
$$$$
-[(\Gg_1^-\Gg_4^-)^2+(\Gg_1^+\Gg_4^+)^2](\cos 2\Gb+3)^2
+2k_0^2\sin^2\Gb\{-u_4^+(3\cos 2\Gb+1)
$$$$
+2\cos^2\Gb[-2\Gg_1^+\Gg_4^+(\cos 2\Gb-3)
+\Gg_1^-\Gg_4^-(\cos 2\Gb+3)]
-4u_4^--2k_0^2\sin^2\Gb(1+\cos^4\Gb)\},
$$$$
h_3=
4 \sin^4\Gb[u_4^--u_4^++2(\Gg_1^-\Gg_4^--\Gg_1^+\Gg_4^+)\cos^2\Gb+2k_0^2\sin^4\Gb],
$$
\beq
u_1=(\Gg_1^+)^2 - k_0^2, \quad u_2=(\Gg_4^+)^2 - k_0^2, \quad u_3= \Gg_1^- \Gg_4^- + k_0^2 \cos^2\Gb, \quad
u_4^\pm=(\Gg_1^\pm)^2+(\Gg_4^\pm)^2.
\label{3.16'}
\eeq

Let $\pm a_j$ ($j=1,2,3,4$) be the eight branch point of the function $\sqrt{f(\Gn)}$ (they are determined explicitly by radicals) such that $0< \I a_1 \le \I a_2 \le \I a_3 \le \I a_4$.
Cut the plane along the segments $l_1^\pm$ and $l_2^\pm$ with the starting points $\pm a_1$ and $\pm a_3$ and the terminal points
$\pm a_2$ and $\pm a_4$, respectively.
Fix the single branch of the  function $\sqrt{f(\Gn)}$ as follows:
\beq
\sqrt{f(\Gn)}=2i\sin^4\Gb\prod_{j=1}^4\sqrt{\Gr^+_j\Gr_j^-}e^{i(\Gvf_j^++\Gvf_j^-)/2},
\label{3.16''}
\eeq
where $\Gr_j^\pm=|\Gn\mp a_j|$, $\Gvf_j^\pm=\arg(\Gn\mp a_j)$, $\Gvf_j^+\in(-\pi,0)$, $\Gvf_j^-\in(0,\pi)$, 
$\Gn\in L$, $j=1,2,3,4$.

The genus-3 hyperelliptic surface $\frak R$ is formed by  two copies, ${\Bbb C}_1$ and ${\Bbb C}_2$, of the extended
$\Gn$-plane ${\Bbb C}\cup\{\infty\}$ cut along the segments $l_1^\pm$ and $l_2^\pm$. The two sheets are glued according to the rule
\beq
w=\left\{\begin{array}{cc}
\sqrt{f(\Gn)}, & \Gn\in \Bbb C_1,\\
-\sqrt{f(\Gn)}, & \Gn\in \Bbb C_2.\\
\end{array}
\right.
\label{3.17}
\eeq 
Let $\Gl(\Gn,w)$ be a new function defined on the surface $\frak R$ as
\beq
\Gl(\Gn,w)= \left\{\begin{array}{cc}
\Gl_1(\Gn), & \Gn\in  \Bbb C_1,\\
\Gl_2(\Gn), & \Gn\in \Bbb C_2,\\
\end{array}
\right.
\label{3.18}
\eeq
where $\Gl_1(\Gn)=b_0(\Gn)+c_0(\Gn)\sqrt{f(\Gn)}$  and $\Gl_1(\Gn)=b_0(\Gn)-c_0(\Gn)\sqrt{f(\Gn)}$ are the eigenvalues of the matrix $\GG(\Gn)$.
We find it convenient to introduce  two matrices, $Y$ and $Q$, as 
\beq
Y(\Gn,w)=\fr12\left[I+\fr{1}{w}Q(\Gn)\right],\quad 
Q(\Gn)=
\left(\begin{array}{cc}
l(\Gn) & m(\Gn)\\
n(\Gn) & -l(\Gn) \\
\end{array}
\right).
\label{3.19}
\eeq
Then the Wiener-Hopf matrix factors of the matrix $\GG(\Gn)$
may be expressed through a nontrivial  solution to the associated scalar RHP on the surface $\frak R$ {\bf(\ref{moi})}, {\bf(\ref{antmoi})}, {\bf(\ref{antsil})}.

{\sl Theorem 4.1. Let $\Gc(\Gn,w)$ be a nontrivial solution to the following RHP.
 
Find a function $\Gc(\Gn,w)$ piece-wise analytic on the surface  $\frak R\setminus\frak L$,   $\frak L=(L\subset  \Bbb C_1)\cup (L\subset \Bbb C_2)$,
except for at most a finite number of poles, H\"older-continuous up to the contour $\frak L$, bounded at both infinite points of the surface $\frak R$ and satisfying the boundary condition 
\beq
\Gc^+(\Gn,w)=\Gl(\Gn,w)\Gc^-(\Gn,w), \quad (\Gn,w)\in\frak L.
\label{3.21}
\eeq
Then the matrix $X(\Gn)$ and  its inverse 
$$
X(\Gn)=\Gc(\Gn,w)Y(\Gn,w)+\Gc(\Gn,-w)Y(\Gn,-w),\quad \Gn\in{\Bbb C^\pm},
$$
\beq
[X(\Gn)]^{-1}=\fr{Y(\Gn,w)}{\Gc(\Gn,w)}+\fr{Y(\Gn,-w)}{\Gc(\Gn,-w)}, \quad \Gn\in{\Bbb C^\pm},
\label{3.22}
\eeq
provide a piece-wise meromorphic  solution of the  matrix factorization problem 
\beq
\GG(\Gn)=X^+(\Gn)[X^-(\Gn)]^{-1}, \quad \Gn\in L.
\label{3.20}
\eeq
}

\subsection{Solution with an essential singularity at infinity}

The chief difficulty in the procedure of matrix factorization based on the use of the solution of the scalar RHP on a Riemann surface (\ref{3.21})
arises in the necessity of constructing a meromorphic solution with explicitly determined poles and zeros. In order to derive such a solution, we consider
the function
\beq
\Gc_0(\Gn,w)=\exp\{\psi_0(\Gn,w)\},\quad (\Gn,w)\in{\frak R},
\label{3.27}
\eeq
where $\psi_0(\Gn,w)$ is the Weierstrass integral
\beq
\psi_0(\Gn,w)=\fr{1}{2\pi i}\int_{\frak L}\log\Gl(t,\Gx)\fr{w+\Gx}{2\Gx}\fr{dt}{t-\Gn},
\label{3.28}
\eeq
and $\Gx=w(t)$, $t\in{\frak L}$. The integral  over the contour $\CL$ on the surface $\frak R$ 
can be transformed into two integrals 
over the contour $L$ in the $\Gn$-plane as
\beq
\psi_0(\Gn,w)=\fr{1}{4\pi i}\int_L [\log\Gl_1(t)+\log\Gl_2(t)]\fr{dt}{t-\Gn}+\fr{w}{4\pi i}\int_L
 [\log\Gl_1(t)-\log\Gl_2(t)]
\fr{dt}{\sqrt{f(t)}(t-\Gn)}.
\label{3.28'}
\eeq
To simplify this representation, we study the eigenvalues $\Gl_j$ of the matrix $\GG(\Gn)$. Referring to  (\ref{3.12}), we obtain that
$b(\Gn)=b(-\Gn)$, $c(\Gn)=-c(-\Gn)$ and also
\beq
\Gl_1(\Gn)\Gl_2(\Gn)=\det\GG(\Gn)=1, \quad \fr{\Gl_1(\Gn)}{\Gl_2(\Gn)}=\fr{b(\Gn)+c(\Gn)\sqrt{f(\Gn)}}{b(\Gn)-c(\Gn)\sqrt{f(\Gn)}}=\fr{\Gl_2(-\Gn)}{\Gl_1(-\Gn)}.
\label{3.23}
\eeq
As $\Gn\to\infty$,
\beq
b(\Gn)=-\fr{i}{\Gd_*(\Gg_1^++\Gg_4^+)}+O(\Gn^{-1}),  \quad c(\Gn)\sim -\fr{\hat\Gg\cos\Gb}{\Gn^5(\Gg_1^++\Gg_4^+)\sin^6\Gb},\quad 
\Gn\to\infty.
\label{3.24}
\eeq
On putting $\Gd_*=-i(\Gg_1^++\Gg_4^+)^{-1}$, we have $b(\Gn)=1+O(\Gn^{-1}),$  $\Gn\to\infty$.  At zero and at infinity, the function $\Gl_1(\Gn)/\Gl_2(\Gn)$
has the following properties:
\beq
 \fr{\Gl_1(0)}{\Gl_2(0)}=1, \quad \fr{\Gl_1(\Gn)}{\Gl_2(\Gn)}=1+O(\Gn^{-1}), \quad \Gn\to\infty.
\label{3.25}
\eeq
It has also been established that $\Gl_1(\Gn)$ and $\Gl_2(\Gn)$ are bounded and do not vanish on the contour $L$.
On fixing the branches of the functions $\log\Gl_1(\Gn)$ and $\Gl_2(\Gn)$ as $\log\Gl_1(\infty)=\log\Gl_2(\infty)=0$ 
we obtain that $\log\Gl_2(\Gn)=-\log\Gl_1(\Gn)$ for all $\Gn\in L$ and therefore
\beq
\psi_0(\Gn,w)=\fr{w}{4\pi i}\int_L\fr{\Gve(t)dt}{\sqrt{f(t)}(t-\Gn)},
\label{3.28''}
\eeq
where
\beq
\Gve(\Gn)=\log l(\Gn), \quad l(\Gn)=\fr{\Gl_1(\Gn)}{\Gl_2(\Gn)},
\label{3.29}
\eeq
and the branch of the logarithmic function is  fixed by the condition $\Gve(\infty)=0$. 
To establish whether the function $\Gve(\Gn)$ is continuous in the contour $L$, we study next the increment of the argument of 
the function $l(\Gn)$
when $\Gn$ traverses the contour $L$ in the positive direction.  Introduce the index of the function $l(\Gn)$
\beq
\ind l(\Gn)=\fr{1}{2\pi}[\arg l(\Gn)]_{-\infty}^\infty.
\label{3.29.1}
\eeq
Due to the fact that $\Gl_2(-\Gn)=\Gl_1(\Gn)$, the function $\Gve(\Gn)$ is odd, and the index of the function $l(\Gn)$ is an even number,
 $\ind l(\Gn)=2\Gk_0$. If $\Gk_0=0$, then the function $\Gve(\Gn)$ is continuous at the point $\Gn=0$, otherwise it is discontinuous at $\Gn=0$. 
 Notice also that
\beq
\Gk_0=\fr{1}{2\pi}[\arg\Gl_1(\Gn)]_{-\infty}^\infty=\fr{1}{2\pi}\left[\arg\fr{\Gl_1(\Gn)}{\Gl_2(\Gn)}\right]_0^\infty.
\label{3.29.2}
\eeq
 
\begin{figure}[t]
\centerline{
\scalebox{0.7}{\includegraphics{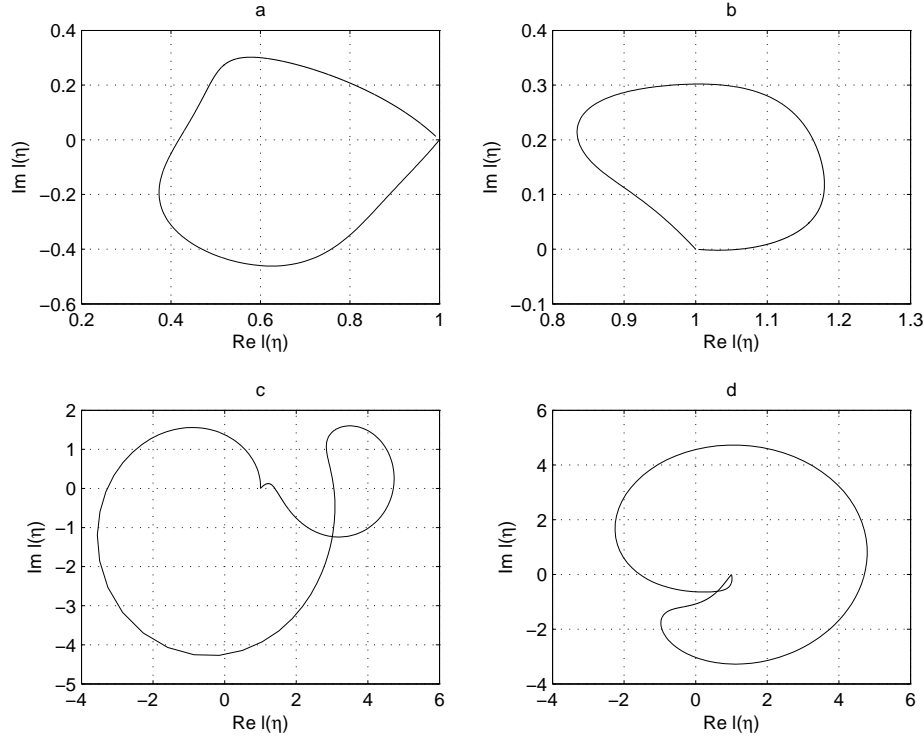}}}
\caption{Parametrically defined function $l(\Gn)$, $0\le\Gn<+\infty$, in the case (i, section 3) when $k_0=1+0.1 i$, $\beta=\pi/4$.
Case a: $\Gk_0=0$ ($\Gg_1^+=1-i$,  $\Gg_4^+=1+2i$, $\Gg_1^-=1-2i$, $\Gg_4^-=1-3i$).
Case b: $\Gk_0=0$ ($\Gg_1^+=2+i$,  $\Gg_4^+=1+2i$, $\Gg_1^-=1-i$, $\Gg_4^-=1-2i$). Case c: $\Gk_0=1$ (the anticlockwise direction, $\Gg_1^+=-1+i$,  $\Gg_4^+=-1+2i$, $\Gg_1^-=-1-0.1i$, $\Gg_4^-=-1-0.2i$). Case d: $\Gk_0=-1$ (the clockwise direction, $\Gg_1^+=1+i$,  $\Gg_4^+=1-i$, $\Gg_1^-=1-0.3i$, $\Gg_4^-=1-i$).}
\label{fig2}
\end{figure}

\begin{figure}[t]
\centerline{
\scalebox{0.7}{\includegraphics{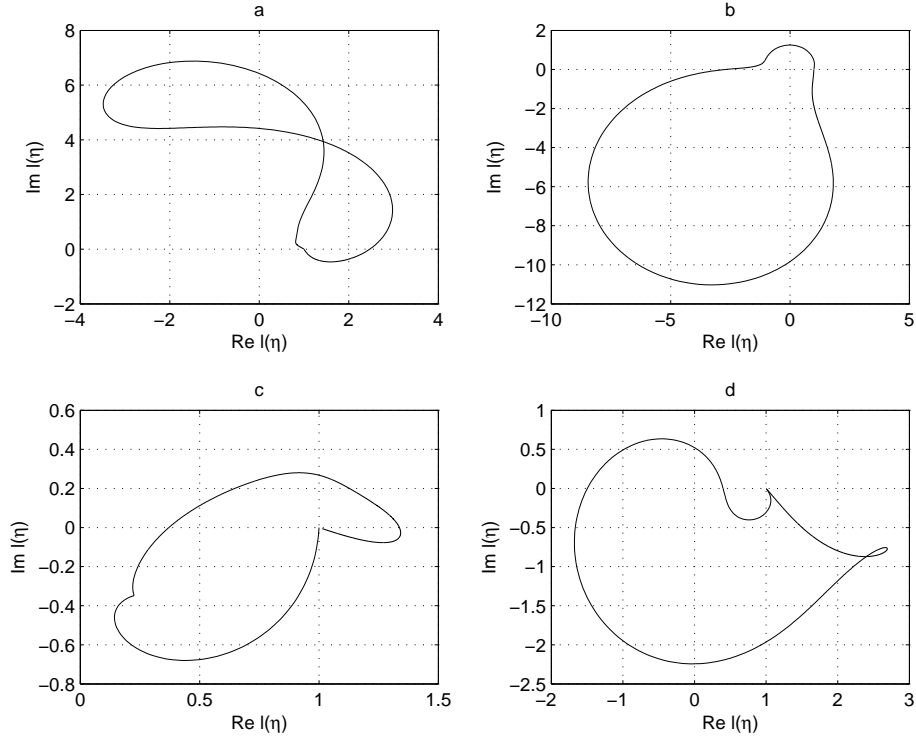}}}
\caption{Parametrically defined function $l(\Gn)$, $0\le\Gn<+\infty$, in cases (ii, section 3): a, b and (iii): c, d, when $k_0=1+0.1 i$, $\beta=\pi/4$.
Case a: $\Gk_0=0$ ($\Gg_1^+=1.5+0.5i$,  $\Gg_4^+=1+i$, $\Gg_1^-=1-0.5i$, $\Gg_4^-=1+i$).
Case b: $\Gk_0=1$ (the anticlockwise direction, $\Gg_1^+=1+i$,  $\Gg_4^+=1-i$, $\Gg_1^-=1-i$, $\Gg_4^-=1+i$). Case c: $\Gk_0=0$ ($\Gg_1^+=1+3i$,  $\Gg_4^+=1+4i$, $\Gg_1^-=1+i$, $\Gg_4^-=1+2i$). Case d: 
$\Gk_0=-1$ (the clockwise direction, $\Gg_1^+=1+0.5i$,  $\Gg_4^+=1+i$, $\Gg_1^-=2+0.5i$, $\Gg_4^-=1+2i$).}
\label{fig3}
\end{figure}

Numerical computations implemented for different sets of the problem parameters $k_0$, $\beta$, $\Gg_1^\pm$
and $\Gg_4^\pm$ show that $\Gk_0$ may be 0, 1, or -1. 
Figure 2  shows samples of the graph of the function $l(\Gn)$ for $0 \le\Gn<+\infty$ in the plane $(\R l(\Gn), \I l(\Gn))$ for some values of the problem parameters in the case (i) when both zeros of the polynomial $\Gd_0(\Gn)$ lie in the lower
half-plane. The graphs of the function $l(\Gn)$ for some values of the problem parameters
in the case (ii), when  the two zeros of $\Gd_0(\Gn)$ lie in the opposite half-planes,
are given in Figures 3a and 3b. Figures 3c and 3d illustrate the case (iii)  (both zeros lie in the upper half-plane) when index $\Gk_0$ vanishes and equals -1, respectively.
In the last case $\Gn$ traverses the contour
around the origin 
in the negative (clock-wise) direction. 

Due to the choice of the branch of the logarithmic function $\log l(\Gn)$, if $\Gk_0=1$, then $\Gve(0^\pm)=\mp 2\pi i$, 
 $\Gve(0^\pm)=\pm 2\pi i$ when $\Gk_0=-1$, and $\Gve(\pm 0)=0$ in the case $\Gk_0=0$. 
Transform now the integral (\ref{3.28''}) 
into the form
\beq
\psi_0(\Gn,w)=\fr{w}{2\pi i}\int_0^\infty\fr{\Gve(t)tdt}{\sqrt{f(t)}(t^2-\Gn^2)}.
\label{3.29.3}
\eeq
This integral is bounded as $\Gn\to 0$ when $\Gk_0=0$ and has a logarithmic singularity otherwise,
\beq
\psi_0(\Gn,w)=(-1)^{j-1}\Gk_0\log\Gn+O(1), \quad \Gn\to 0, \quad (\Gn,w)\in{\Bbb C}_j,\quad j=1,2.
\label{3.29.4}
\eeq
It can  be directly verified that  the function
\beq
\hat\psi_0(\Gn,w)=\psi_0(\Gn,w)+\Gk_0\sum_{j=1}^2(-1)^{j-1}\int_{q_{0j}}^{q_{1j}}\fr{w+\Gx}{2\Gx}\fr{dt}{t-\Gn}
\label{3.29.5}
\eeq
is bounded at the point $\Gn=0$. Here, 
$q_{0j}=(0,(-1)^{j-1}\sqrt{f(0)})\in{\Bbb C_j}$
are the two zero points of the surface,  and $q_{1j}=((-1)^{j-1}\Gr_0, (-1)^{j-1}\sqrt{f(\Gr_0)})\in{\Bbb C_j}$, where
$\Gr_0$ is an arbitrary fixed point in the upper half-plane which coincides with none of the branch points $a_j$ ($j=1,2,3,4$), the point $\Gn_0$,
and the roots of the polynomials $d_1(\Gn)$ 
and $\Gd_0(\Gn)\Gd_0(-\Gn)$ (the final solution of the vector RHP is invariant with respect to the choice of the point $\Gr_0$).  
Because of the symmetry of the points $q_{11}$ and $q_{12}$ we can rewrite the
function $\hat\psi_0(\Gn,w)$ as
\beq
\hat\psi_0(\Gn,w)=\psi_0(\Gn,w)+\Gk_0\left[\int_0^{\Gr_0}\fr{\Gn dt}{t^2-\Gn^2}+w\int_0^{\Gr_0}\fr{tdt}{\sqrt{f(t)}(t^2-\Gn^2)}\right].
\label{3.29.6}
\eeq
\begin{figure}[t]
\centerline{
\scalebox{0.6}{\includegraphics{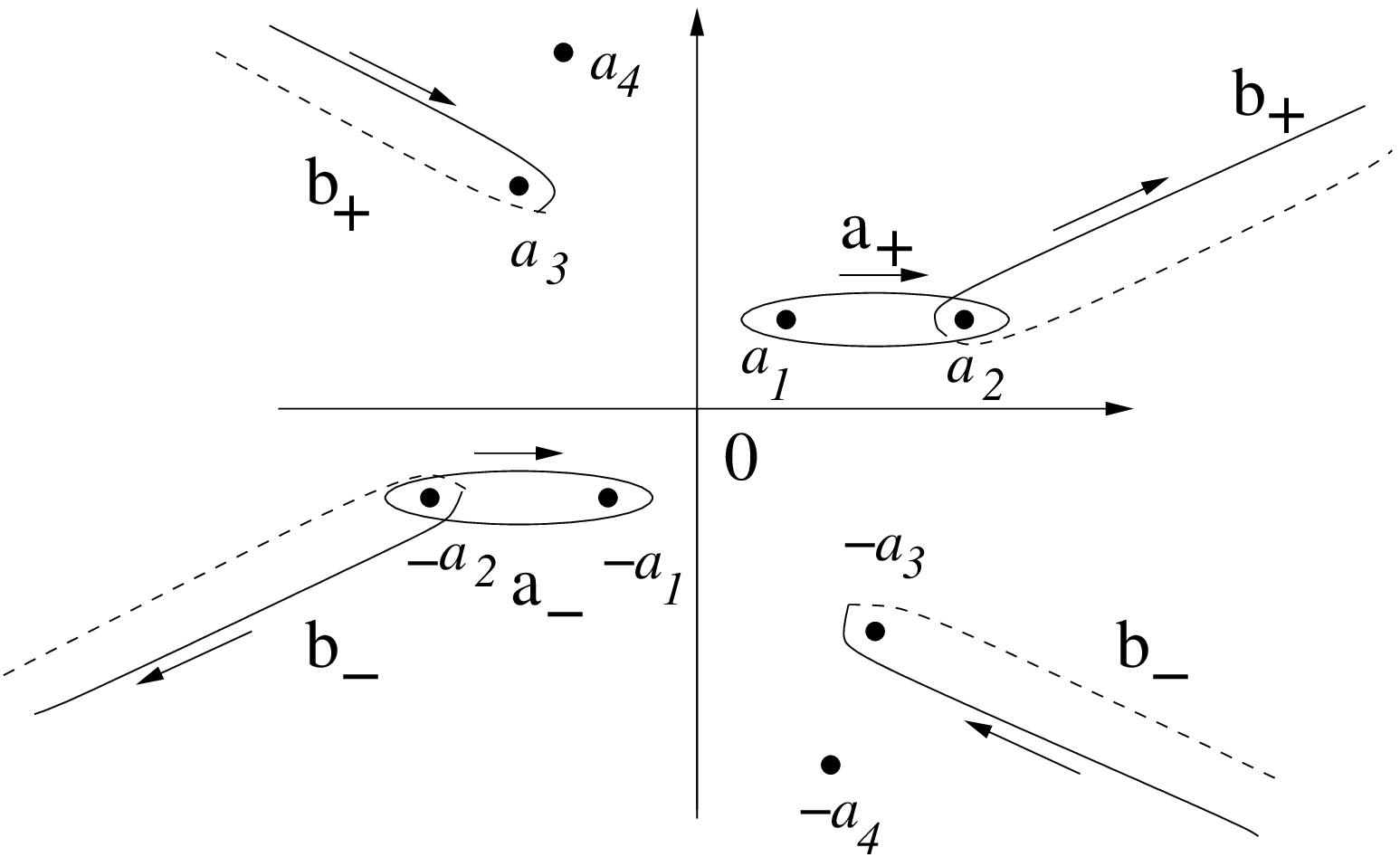}}}
\caption{The loops $\Ba_\pm$ and $\Bb_\pm$.}
\label{fig4}
\end{figure} 

The function $\exp\{\hat\psi_0(\Gn,w\}$ satisfies the boundary condition of the RHP (\ref{3.21}). However,
at the infinite points of the surface it has essential singularities due to the second order pole at infinity of the function (\ref{3.29.6}). 
In order to remove this singularity, we cut the surface $\frak R$ along two pairs of loops $\Ba_-$,  $\Bb_-$
and $\Ba_+$,  $\Bb_+$ (Fig. 4). The loops $\Ba_\pm$ intersect the loops $\Bb_\pm$ at the points  $\pm a_2$, respectively. 
The cross-sections $\Ba_-$ and $\Ba_+$ and $\Bb_-$ and $\Bb_+$ are symmetric
with respect to the origin. The curves $\Ba_-$ and $\Ba_+$  are closed contours which join the branch points $-a_1$ with $-a_2$ and $a_1$ with $a_2$, 
respectively, and lie on both sheets of the surface. The loops  $\Bb_\pm$  join the points $\pm a_2$ with $\pm a_3$ and pass through the infinite points,
where they touch each other but do no intersect. Both parts of the loop $\Bb_-$ lie on the lower half-planes of both sheets, while
the contour $\Bb_+$ lies on the upper half-planes.
The solid lines in Fig. 4 correspond
to the parts of the loops $\Bb_\pm$ on the sheet $\Bbb C_1$, while the broken lines show the parts of the loops on the second sheet.
Notice that the system of loops $\Ba_\pm$, $\Bb_\pm$ does not form a system of canonical cross-section of the genus-3 surface $\frak R$
which comprises three pairs of loops. Due to the special symmetry of the problem,
 to remove the essential singularity at infinity,  it suffices to use the two cross-sections $\Ba_\pm$ and  $\Bb_\pm$. Analyze now the function
\beq
\tilde\psi_0(\Gn,w)=
\left[\int_{c_-}+\int_{c_+}+m_0\left(\oint_{\Ba_-}+\oint_{\Ba_+}\right)+n_0\left(\oint_{\Bb_-}+\oint_{\Bb_+}\right)\right]\fr{w+\Gx}{2\Gx}\fr{dt}{t-\Gn}. 
\label{2.29.7}
\eeq 
Here, the contours $c_-$ and $c_+$
are smooth simple curves on the surface which  do not cross the loops $\Ba_\pm$ and $\Bb_\pm$. The starting and terminal points
of the contours $c_\pm$ are $\pm p_0$ and $\pm p_1$, respectively.
The point $p_0=(\Gs_0,\sqrt{f(\Gs_0)})$ is an arbitrary fixed point in the sheet ${\Bbb C_1}$ (the final solution of the vector RHP is independent of this point),
while the point $p_1=(\Gs_1, w(\Gs_1))$ may lie on either sheet of the surface and has to be determined.
The numbers $m_0$ and   $n_0$
are integers to be fixed.   The function $\exp\{\tilde\psi_0(\Gn,w)\}$ is meromorphic
in any finite part of the surface $\frak R$ and continues through the contours $c_\pm$, $\Ba_\pm$ and $\Bb_\pm$. 
Add the function $\tilde\psi_0(\Gn,w)$ to $\hat\psi_0(\Gn,w)$.  A simple alteration of this sum, $\psi(\Gn,w)$, implies 
$$
\psi(\Gn,w)=\fr{w}{2\pi i}\int_0^\infty\fr{\Gve(t)tdt}{\sqrt{f(t)}(t^2-\Gn^2)}+\Gk_0\int_0^{\Gr_0}
\left(\Gn+\fr{wt}{\sqrt{f(t)}}
\right)
\fr{dt}{t^2-\Gn^2}
$$
\beq
+\left(\int_{p_0}^{p_1}+m_0\oint_{\Ba_+}+n_0\oint_{\Bb_+}\right)\fr{w+\Gx}{\Gx}\fr{tdt}{t^2-\Gn^2}.
\label{3.31}
\eeq 
The function $\Gc(\Gn,w)=\exp\{\psi(\Gn,w)\}$ is meromorphic  in each finite part of the Riemann surface and
satisfies the boundary condition  (\ref{3.21}). However, due to the pole of the Weierstrass kernel at infinity,
it has essential singularities at both infinite points of the surface, $(\infty,\infty_1)$ and $(\infty,\infty_2)$.

\subsection{Jacobi inversion problem}

The analysis of the solution at the points $(\infty,\infty_1)$ and $(\infty,\infty_2)$ shows that both functions, $\psi(\Gn,w)$ and
$\Gc(\Gn,w)=\exp\{\psi(\Gn,w)\}$, are  bounded as $\Gn\to\infty$ if and only if
the following condition holds
\beq
\fr{1}{2\pi i}\int_0^\infty\fr{\Gve(t)tdt}{\sqrt{f(t)}}
+\Gk_0\int_0^{\Gr_0}\fr{tdt}{\sqrt{f(t)}}
+\int_{p_0}^{p_1}\fr{tdt}{\Gx}+m_0\oint_{\Ba_+}\fr{tdt}{\Gx}+n_0\oint_{\Bb_+}\fr{tdt}{\Gx}=0.
\label{3.34}
\eeq
In this section we aim to reduce the condition (\ref{3.34}) to a genus-1 Jacobi inversion problem 
and solve it by inversion of an elliptic integral.
Generically,   a scalar RHP on a genus-$\Gr$  surface requires solving the associated Jacobi inversion problem in terms of the zeros
 of the genus-$\Gr$ Riemann $\Gt$-function  {\bf(\ref{zve})}. However, due to the symmetry properties of the problem, 
 the genus of the  $\Gt$-function can be decreased. In our case we have managed to avoid the genus-3 $\Gt$-function associated with the surface
 $\frak R$ and decrease it genus to $\Gr=1$ because
 
 (i) the degree-8 characteristic polynomial $f(\Gn)$ is even and
 
 (ii) the function $\Gve(\Gn)$ is odd.

To solve the nonlinear equation (\ref{3.34}) that is to define the point $p_1\in{\frak R}$ and
the integers $m_0$ and $n_0$,  
first, we express the integrals over the loops  $\Ba_+$ and $\Bb_+$ 
$$
\oint_{\Ba_+}\fr{tdt}{\Gx}=\int_{a_1^2}^{a_2^2}\fr{dt_*}{\sqrt{f_*(t_*)}}, 
$$
\beq
\oint_{\Bb_+}\fr{tdt}{\Gx}=\int_{a_2^2}^\infty\fr{dt_*}{\sqrt{f_*(t_*)}}+\int_{\infty}^{a_3^2}\fr{dt_*}{\sqrt{f_*(t_*)}}
\label{3.34'}
\eeq
through elliptic integrals. Here, $f(t)=f_*(t^2)$, and
the branch $\sqrt{f_*(\Gn_*)}$ is  fixed by the condition $\sqrt{f_*(\Gn_*)}\sim 2i\Gn_*^2\sin^4\Gb$, $\Gn_*\to\infty$,
in the $\Gn_*$-plane ($\Gn_*=\Gn^2$) cut along the two lines joining the branch points, $a_1^2$ with $a_2^2$ and
$a_3^2$ with $a_4^2$. 
 By making the homographic transformation {\bf(\ref{han})}
$t_*=(\Gb_1+\Gb_2\tau)(1-\mu\tau)^{-1}$ we fix the parameters $\Gb_1$, $\Gb_2$ and $\mu$
such that the points $a_1^2$, $a_2^2$, $a_3^2$, and $a_4^2$
are mapped into the points $1$, $1/\Gk$, $-1/\Gk$ and  $-1$, respectively  ($\Gk$ is to be determined).
 It directly follows that
\beq
\fr{t_*-a_4^2}{a_1^2-a_4^2}=\fr{(1-\mu)(1+\tau)}{2(1-\mu\tau)}, \quad \fr{t_*-a_1^2}{a_4^2-a_1^2}=\fr{(1+\mu)(1-\tau)}{2(1-\mu\tau)},
\label{3.35}
\eeq
and
\beq
\fr{t_*-a_3^2}{a_2^2-a_3^2}=\fr{(1-\mu/\Gk)(1+\Gk\tau)}{2(1-\mu\tau)}, \quad \fr{t_*-a_2^2}{a_3^2-a_2^2}=\fr{(1+\mu/\Gk)(1-\Gk\tau)}{2(1-\mu\tau)}.
\label{3.36}
\eeq
Equations (\ref{3.35}) imply
\beq
\fr{t_*-a_4^2}{t_*-a_1^2}=\fr{(\mu-1)(1+\tau)}{(\mu+1)(1-\tau)}.
\label{3.37}
\eeq
On putting in this equation $t_*=a_2^2$, $\tau=1/\Gk$ and  $t_*=a_3^2$, $\tau=-1/\Gk$ and denoting
 $\mu_*=(\mu-1)/(\mu+1)$, $\Gk_*=(\Gk-1)/(\Gk+1)$  we obtain  the system of equations 
\beq
\fr{\mu_*}{\Gk_*}=\fr{a_2^2-a_4^2}{a_2^2-a_1^2}, \quad \mu_*\Gk_*=\fr{a_3^2-a_4^2}{a_3^2-a_1^2}.
\label{3.38}
\eeq
It has two sets of solutions defined by
\beq
\Gk_*^2=\fr{(a_2^2-a_1^2)(a_3^2-a_4^2)}{(a_2^2-a_4^2)(a_3^2-a_1^2)},\quad \mu_*=\fr{(a_2^2-a_4^2)\Gk_*}{a_2^2-a_1^2}.
\label{3.39}
\eeq
Each of them determines the parameters $\mu$ and $\Gk$,
\beq
\mu=\fr{1+\mu_*}{1-\mu_*}, \quad \Gk=\fr{1+\Gk_*}{1-\Gk_*}.
\label{3.40}
\eeq
By expressing $t_*$ from the two relations (\ref{3.35}) and adding the formulas obtained we have the homographic transformation
sought
\beq
t_*=\fr{a_1^2+a_4^2}{2}+\fr{(a_1^2-a_4^2)(\tau-\mu)}{2(1-\mu\tau)}.
\label{3.41}
\eeq
We next differentiate the two equations (\ref{3.36})
and find
\beq
dt_*=\fr{\sqrt{(1-\mu^2)(\Gk-\mu^2/\Gk)(a_2^2-a_3^2)(a_1^2-a_4^2)}}{2(1-\mu\tau)^2}d\tau.
\label{3.42}
\eeq
On multiplying the four relations in (\ref{3.35}) and (\ref{3.36}) and using (\ref{3.42}) we deduce
\beq
\int\fr{dt_*}{\sqrt{f_*(t_*)}}=h\int\fr{d\tau}{\sqrt{(1-\tau^2)(1-\Gk^2\tau^2)}},
\label{3.43}
\eeq
where
\beq
h=\fr{\Ge}{i\sin^4\Gb}\sqrt{\fr{\Gk}{(a_2^2-a_3^2)(a_1^2-a_4^2)}},\quad \Ge=\pm 1,
\label{3.44}
\eeq
and $v(\tau)=\sqrt{(1-\tau^2)(1-\Gk^2\tau^2)}$ is the branch fixed by the condition $v(0)=1$ of the two-valued function $v^2=(1-\tau^2)(1-\Gk^2\tau^2)$
in the $\tau$-plane cut along the segments
$\hat l_1$ and $\hat l_2$ with the starting and
terminal points $-1/\Gk$ and $-1$ for  $\hat l_1$ and $1$ and $1/\Gk$  for $\hat l_2$. 
We can now compute the  integrals over the loops $\Ba_+$ and $\Bb_+$ in (\ref{3.34'})
$$
 \oint_{\Ba_+}\fr{tdt}{\Gx}= h\int_1^{1/\Gk}\fr{d\tau}{v(\tau)}=  hi{\BK}',
 $$
\beq
\oint_{\Bb_+}\fr{tdt}{\Gx}=2h\int_{1/\Gk}^\infty\fr{d\tau}{v(\tau)}=-2h\BK,
\label{3.46}
\eeq
where $\BK'=\BK(\sqrt{1-\Gk^2})$, and $\BK=\BK(\Gk)$ is the complete elliptic integral  of the first kind.
Denote next the inverse to (\ref{3.41}) transformation as
\beq
\tau=u(t_*)=\fr{t_*-a_4^2-\mu_*(t_*-a_1^2)}{t_*-a_4^2+\mu_*(t_*-a_1^2)}
\label{3.49'}
\eeq
and let $\hat\Gs_j=u(\Gs_j^2)$, $j=0,1$.
On making this transformation in the third integral in (\ref{3.34}) we derive
\beq
\int_{p_0}^{ p_1}\fr{tdt}{\Gx}=\fr{h}{2}\left(\int_{\hat\Gs_0}^0\fr{d\tau}{v(\tau)}+\int_0^{\hat\Gs_1}\fr{d\tau}{v(\tau)}
\right)
\label{3.48}
\eeq
in the case $p_1\in{\Bbb C}_1$ and
\beq
\int_{p_0}^{p_1}\fr{tdt}{\Gx}=\fr{h}{2}\left(\int_{\hat\Gs_0}^{0}\fr{d\tau}{v(\tau)}-2\int_{0}^{1}\fr{d\tau}{v(\tau)}
-\int_0^{\hat\Gs_1}\fr{d\tau}{v(\tau)}
\right)
\label{3.49}
\eeq
in the case $p_1\in{\Bbb C}_2$.  The transformation $u$ maps the points $0$, $\infty$ and $\Gr_0^2$ of the $t_*$-plane into the points
$u_0=(a_4^2-\mu_* a_1^2)(a_4^2+\mu_* a_1^2)^{-1}$, $1/\mu$ and $u_1=u(\Gr_0^2)$ of the $\tau$-plane, respectively.
It is helpful to denote
\beq
\hat d=-\fr{1}{2\pi i}\int_{u_0}^{1/\mu}\fr{\hat\Gve(\tau)d\tau}{v(\tau)}
-\Gk_0\int_{u_0}^{u_1}
\fr{d\tau}{v(\tau)}+
\int_0^{\hat\Gs_0}\fr{d\tau}{v(\tau)}, \quad \hat\Gve(\tau)=\Gve(\sqrt{t_*}).
\label{3.51}
\eeq
Now the Jacobi inversion problem (\ref{3.34}) may be put in a more convenient form. Let first  $p_1\in{\Bbb C}_1$. Then
\beq
\int_0^{\hat\Gs_1}\fr{d\tau}{v(\tau)}=\hat d+4n_0\BK-2im_0\BK'.
\label{3.50}
\eeq
By inversion of 
the elliptic integral in (\ref{3.50}) we find $\hat\Gs_1={\rm sn}\, \hat d$ and therefore 
the affix $\Gs_1$ of the point $p_1$ is given by
\beq
\Gs_1=\pm\sqrt{\fr{a_1^2+a_4^2}{2}+\fr{(a_1^2-a_4^2)(\hat\Gs_1-\mu)}{2(1-\mu\hat\Gs_1)}}.
\label{3.52}
\eeq
Because of the symmetry, either sign leads to the same solution. 
The integers $m_0$ and $n_0$ can be directly found from (\ref{3.50}) 
\beq
m_0=-\fr{\I[(I_*-\hat d)\ov{\BK}]}{2\R[\BK\ov{\BK'}]}, \quad
n_0=\fr{\R[(I_*-\hat d)\ov{\BK'}]}{4\R[\BK\ov{\BK'}]},
\label{3.53}
\eeq
where $I_*$ is an elliptic integral of the first kind
\beq
I_*=\int_0^{\hat\Gs_1}\fr{d\tau}{v(\tau)}=F(\sin^{-1}({\rm sn}\, \hat d),\Gk).
\label{3.54}
\eeq
If at least one of the numbers $m_0$, $n_0$ is not integer, then $p_1\in{\Bbb C}_2$, and
 $n_0$, $m_0$ are both necessarily integer. In this case the Jacobi inversion problem (\ref{3.34}) becomes
\beq
-\int_0^{\hat\Gs_1}\fr{d\tau}{v(\tau)}=\hat d+4\left(n_0+\fr12\right)\BK-2im_0\BK'.
\label{3.55}
\eeq 
The affix of the point $p_1$ is the same as in the previous case, while the integers $m_0$ and $n_0$ are
different,
\beq
m_0=\fr{\I[(I_*+\hat d)\ov{\BK}]}{2\R[\BK\ov{\BK'}]}, \quad
n_0=-\fr12-\fr{\R[(I_*+\hat d)\ov{\BK'}]}{4\R[\BK\ov{\BK'}]}.
\label{3.56}
\eeq
This completes the solution of the Jacobi problem.
Summarize the results of Section 4.

\vspace{.1in}

{\sl Theorem 4.2.  Let $p_0=(\Gs_0,\sqrt{f(\Gs_0)})\in{\Bbb C_1}$, $\Gr_0\in{\Bbb C^+}$ be arbitrary fixed points and none of them coincide with the branch
points of $f^{1/2}(\Gn)$. Let $\Gk_0$ be the integer (\ref{3.29.2}) and $\Gs_1$ be either complex number in (\ref{3.52}).  Choose
the point $p_1$ as $(\Gs_1,\sqrt{f(\Gs_1)})\in{\Bbb C_1}$ if both numbers in (\ref{3.53}) are integers and as $(\Gs_1,-\sqrt{f(\Gs_1)})\in{\Bbb C_2}$
otherwise (in this case both numbers in (\ref{3.56}) are necessarily integers). Define $\Gy(\Gn)$  by (\ref{3.31}). Then the function
\beq
\Gc(\Gn,w)=e^{\Gy(\Gn,w)}, \quad (\Gn,w)\in\frak R,
\label{3.56'}
\eeq
is bounded at the two infinite points of the surface $\frak R$, piece-wise meromorphic on  $\frak R$ with the discontinuity line $\frak L$ in which it
 satisfies the boundary condition  (\ref{3.21}) of the scalar RHP.
}

\setcounter{equation}{0}

\section{Solution to the  vector RHP}

\subsection{Matrix factorization and its analysis}

As a matter of utility, it will be desirable to have the solution to the RHP (\ref{3.21}) on the Riemann surface ${\frak R}$
expressed in terms of two functions defined on the $\Gn$-plane
\beq
\Gc(\Gn,w)=\exp\{\psi_1(\Gn)+w\psi_2(\Gn)\},
\label{3.57}
\eeq
where
$$
\psi_1(\Gn)=\fr{\Gk_0}{2}\log\fr{\Gn-\Gr_0}{\Gn+\Gr_0}+
\fr12\log\fr{\Gn^2-\Gs_1^2}{\Gn^2-\Gs_0^2},
$$
$$
\psi_2(\Gn)=\fr{1}{2\pi i}\int_0^\infty\fr{\Gve(t)tdt}{\sqrt{f(t)}(t^2-\Gn^2)}+
\Gk_0\int_0^{\Gr_0}\fr{tdt}{\sqrt{f(t)}
(t^2-\Gn^2)}
$$
\beq
+\left(\int_{p_0}^{p_1}+m_0\oint_{\Ba_+}+n_0\oint_{\Bb_+}\right)\fr{tdt}{\xi(t^2-\Gn^2)},
\label{3.58}
\eeq
 Because the solution to the Jacobi problem, the point $\Gs_1$ and the integers $m_0$ and $n_0$,
satisfy the condition (\ref{3.34}), it is possible to alter  formula (\ref{3.58}) for the function $\psi_2(\Gn)$ as
$$
\psi_2(\Gn)=\fr{1}{\Gn^2}\left\{\fr{1}{2\pi i}\int_0^\infty\fr{\Gve(t)t^3dt}{\sqrt{f(t)}(t^2-\Gn^2)}+
\Gk_0\int_0^{\Gr_0}\fr{t^3dt}{\sqrt{f(t)}
(t^2-\Gn^2)}
\right.
$$
\beq
\left.
+\left(\int_{p_0}^{p_1}+m_0\oint_{\Ba_+}+n_0\oint_{\Bb_+}\right)\fr{t^3dt}{\xi(t^2-\Gn^2)}
\right\}
\label{3.59}
\eeq
which can  be conveniently used for large $\Gn$. The analysis of this formula shows that $\psi_1(\Gn)+w\psi_2(\Gn)$ is bounded as $\Gn\to\infty$,
and therefore the essential singularity of the solution $\Gc(\Gn,w)$ has been removed.

With the solution to the  RHP on the surface ${\frak R}$ at hand, we may write down formulas (\ref{3.22})
for the factor-matrices $X^\pm(\Gn)$ and their inverses in terms of functions defined
on the $\Gn$-plane 
$$
X^\pm(\Gn)=e^{\psi_1^\pm(\Gn)}\left[
\cosh(f^{1/2}(\Gn)\psi_2^\pm(\Gn))I+\fr{1}{f^{1/2}(\Gn)}\sinh(f^{1/2}(\Gn)\psi_2^\pm(\Gn))Q(\Gn)\right],
$$
\beq
[X^\pm(\Gn)]^{-1}=e^{-\psi_1^\pm(\Gn)}\left[
\cosh(f^{1/2}(\Gn)\psi_2^\pm(\Gn))I-\fr{1}{f^{1/2}(\Gn)}\sinh(f^{1/2}(\Gn)\psi_2^\pm(\Gn))Q(\Gn)\right], \quad \Gn\in{\Bbb C}^\pm.
\label{3.60}
\eeq
It is essential for the solution of the vector RHP to study the behavior of the matrix $X(\Gn)=X^\pm(\Gn)$,
$\Gn\in{\Bbb C}^\pm$, at the points $p_0=(\Gs_0,\sqrt{f(\Gs_0)})\in{\Bbb C}_1$ and $p_1=(\Gs_1,w(\Gs_1))\in\frak R$.
For this purpose we use the representation (\ref{3.22}). From (\ref{3.57}), (\ref{3.58}),
$$
\Gc(\Gn,w)\sim\fr{E^\pm_{01}}{\Gn\mp\Gs_0}, \quad \Gn\to\pm\Gs_0, \quad (\Gn,w)\in{\Bbb C}_1,
$$
\beq
\Gc(\Gn,w)\sim E^\pm_{02}, \quad \Gn\to\pm\Gs_0, \quad (\Gn,w)\in{\Bbb C}_2,
\label{3.60.1}
\eeq
where $E^\pm_{0j}$ are nonzero constants. This yields
\beq
X(\Gn)\sim \fr{E^\pm_{01}}{\Gn\mp\Gs_0}Y^\pm,\quad \Gn\to\pm\Gs_0,
\label{3.60.2}
\eeq
where $Y^\pm=Y(\pm\Gs_0,\sqrt{f(\Gs_0)})$ are rank-1 $2\times 2$
matrices. 

If $p_1$ is  a point of the first sheet of the surface
$\frak R$, then
\beq
\Gc(\Gn,w)\sim E^\pm_{11}(\Gn\mp\Gs_1), \quad \Gc(\Gn,-w)\sim E^\pm_{12},\quad
\Gn\to\pm\Gs_1, \quad (\Gn,  w)\in{\Bbb C}_1.
\label{3.60.3}
\eeq
 In the case $p_1\in{\Bbb C}_2$,  
\beq
\Gc(\Gn,w)\sim E^\pm_{21}, \quad \Gc(\Gn,-w)\sim E^\pm_{22}(\Gn\mp\Gs_1),\quad
\Gn\to\pm\Gs_1, \quad (\Gn,  w)\in{\Bbb C}_1.
\label{3.60.4}
\eeq
Here, $E^\pm_{sj}$ ($s,j=1,2$) are nonzero constants. Then, regardless of whether the point $p_1$ belongs to the first sheet or the second one,
 the inverse matrix $[X(\Gn)]^{-1}$ has poles at the points $\pm\Gs_1$.
 Assume now that $z(\Gn)$ is an order-2 vector whose components have certain nonzero limits at the points $\Gn=\pm\Gs_1$. Then since
$Y(\Gn,w)$ is a rank-1 matrix, we have
 \beq 
[X(\Gn)]^{-1}z(\Gn)\sim\fr{E_j^\pm}{\Gn\mp\Gs_1}\left(\begin{array}{c}
1 \\ s_j^\pm\\
\end{array}\right),\quad \Gn\to\pm\Gs_1, \quad p_1\in{\Bbb C_j},\quad j=1,2,
\label{3.60.5}
\eeq
where $E_j^\pm=$const,  $s_j^\pm=n(\pm\Gs_1)[l(\pm\Gs_1)+w_j]^{-1}$,  and $w_j=(-1)^{j-1}\sqrt{f(\Gs_1)}$
 for $p_1\in{\Bbb C}_j$, $j=1,2$. 
 At the same time, if $\hat z(\Gn)$ is an order-2 vector such that
\beq
\hat z(\Gn)\sim \fr{\hat E_j^\pm}{\Gn\mp\Gs_1}\left(\begin{array}{c}
1 \\ s_j^\pm\\
\end{array}\right),\quad \Gn\to\pm\Gs_1,
\label{3.60.6}
\eeq
where $\hat E_j^\pm$ are nonzero constants, then
\beq
X(\Gn)\hat z(\Gn)=Y(\pm\Gs_1, -w_j)\fr{\hat E_j^\pm}{\Gn\mp\Gs_1}\left(\begin{array}{c}
1 \\ s_j^\pm\\
\end{array}\right)+\tilde z(\Gn)=\tilde z(\Gn), \quad \Gn\to\pm\Gs_1,
\label{3.60.7}
\eeq
where $\tilde z(\Gn)$ is an order-2 vector bounded as $\Gn\to\pm\Gs_1$.

We next examine the matrix $X(\Gn)$ and its inverse as $\Gn\to\pm\Gr_0$. If $\Gk_0=0$, then both functions, $\psi_1(\Gn)$ and $\psi_2(\Gn)$, are bounded at the points $\pm\Gr_0$,
and therefore the matrices $X(\Gn)$   and $[X(\Gn)]^{-1}$ are  bounded as well.
If $|\Gk_0|=1$, then
$$
\Gc(\Gn,w)\sim D_1^+(\Gn-\Gr_0)^{\Gk_0}, \quad \Gn\to\Gr_0, \quad (\Gn,w)\in{\Bbb C_1},
$$
\beq
\Gc(\Gn,w)\sim D_2^+, \quad \Gn\to\Gr_0, \quad (\Gn,w)\in{\Bbb C_2}.
\label{3.60.8}
\eeq
 At the points with affix $-\Gr_0$, 
$$
\Gc(\Gn,w)\sim D_1^-, \quad \Gn\to-\Gr_0, \quad (\Gn,w)\in{\Bbb C_1},
$$
\beq
\Gc(\Gn,w)\sim D_2^-(\Gn+\Gr_0)^{-\Gk_0}, \quad \Gn\to-\Gr_0, \quad (\Gn,w)\in{\Bbb C_2}.
\label{3.60.8'}
\eeq
Here,  $D_j^\pm$ are nonzero constants.
At the points $\Gn=\pm\Gk_0\Gr_0$ ($\Gk_0=\pm 1$), the matrix $X(\Gn)$ and its inverse accordingly behave
$$
X(\Gn)=O(1), \quad 
[X(\Gn)]^{-1}\sim D_\pm'(\Gn-\Gk_0\Gr_0)^{-1}\tilde Y^+_{\Gk_0}, \quad \Gn\to\Gk_0\Gr_0,
$$
\beq
X(\Gn)\sim  D_\pm''(\Gn+\Gk_0\Gr_0)^{-1}\tilde Y^-_{\Gk_0},
\quad [X(\Gn)]^{-1}=O(1), \quad \Gn\to-\Gk_0\Gr_0,
\label{3.60.10}
\eeq
where $\tilde Y_{\Gk_0}^\pm=Y(\pm\Gk_0\Gr_0,\pm\Gk_0\sqrt{f(\Gr_0)})$  are rank-1 $2\times 2$ matrices, and $D_\pm'$, $D_\pm''$ are nonzero constant.

Before proceeding with the solution of the vector RHP, we need to factorize the  function $\sqrt{\GD(\Gn)}$ with  $\GD(\Gn)$ being a rational
function given  by (\ref{3.12''}). Since the function $\GD(\Gn)$ admits the splitting
\beq
 \GD(\Gn)=\fr{(\Gn^2-\tau_1^2)(\Gn^2-\tau_2^2)}{(\Gn^2-t_1^2)(\Gn^2-t_2^2)},
 \label{3.74}
 \eeq
where $\I t_j>0$, $\I \tau_j>0$, $j=1,2$, we immediately obtain
 \beq
 \GD(\Gn)=\fr{\Gr^+(\Gn)}{\Gr^-(\Gn)}, \quad \Gr^\pm(\Gn)=\left\{\fr{(\Gn\pm\tau_1)(\Gn\pm\tau_2)}{(\Gn\pm t_1)(\Gn\pm t_2)}\right\}^{\pm1/2}, 
  \quad \Gn\in{\Bbb C^\pm}.
\label{3.78} 
 \eeq
The branches of the functions $\Gr^\pm(\Gn)$ are chosen such that $\Gr^\pm(\Gn)\to 1$, $\Gn\to\infty$, and the branch cuts join
the branch points $\mp t_1$ with $\mp t_2$ and $\mp \tau_1$ with $\mp \tau_2$.
Referring now to the representations (\ref{3.11}) and (\ref{3.20}) we split the matrix $G(\Gn)$ as
\beq
G(\Gn)=\fr{\Gd_* G^1(\Gn)}{\Gd_0(\Gn)}\Gr^+(\Gn)X^+(\Gn)[\Gr^-(\Gn)X^-(\Gn)]^{-1}, \quad \Gn\in L.
\label{3.63}
\eeq
We remind that $\Gd_0(\Gn)$ and $G^1(\Gn)$ are the polynomial and the polynomial matrix given by (\ref{3.7''''}) and (\ref{3.10}), respectively.

\subsection{Vectors $\GF^+(\Gn)$ and $\GF^-(\Gn)$}

To determine the vectors $\GF^+(\Gn)$ and $\GF^-(\Gn)$, we insert the splitting (\ref{3.63}) into the boundary condition (\ref{3.6}) of the vector RHP and use the formulas
\beq
\hat G^1(\Gn)=d_1(\Gn)[G^1(\Gn)]^{-1},\quad 
\hat G^1(\Gn)=\left(\begin{array}{cc}
g_{22}^1(\Gn) \; & -g_{12}^1(\Gn) \\
-g_{21}^1(\Gn) \; & g_{11}^1(\Gn) \\
\end{array}
\right),
\label{3.64}
\eeq
where $g_{sj}^1$ ($s,j=1,2$) are given by (\ref{3.10}).
On factorizing the polynomial $d_1(\Gn)$, we discover
\beq
\fr{\Gd_0(\Gn)[X^+(\Gn)]^{-1}\hat G^1(\Gn)\GF^+(\Gn)}{\Gd_*\Ga_2(\Gn+t_1)(\Gn+t_2)\Gr^+(\Gn)}=\fr{(\Gn-t_1)(\Gn-t_2)}{\Gr^-(\Gn)}[X^-(\Gn)]^{-1}\GF^-(\Gn), \quad \Gn\in L.
\label{3.65}
\eeq
Here, $\pm t_1$ and $\pm t_2$ are the four zeros of the polynomial  
$d_1(\Gn)=\Ga_2(\Gn^2-t_1^2)(\Gn^2-t_2^2)$, and $\I t_j>0$, $j=1,2$.

We begin with the case (i) when the two zeros of the quadratic polynomial $\Gd_0(\Gn)$ lie in the lower half-plane, $\Gd_0(\Gn)=-(\Gn+\tau_1)(\Gn+\tau_2)\sin^2\Gb$, $\I\tau_j>0$, $j=1,2$.
On applying the principle of analytic continuation and the Liouville theorem, we have
\beq
\fr{\nu (\Gn+\tau_1)(\Gn+\tau_2)[X^+(\Gn)]^{-1}\hat G^1(\Gn)\GF^+(\Gn)}{(\Gn+t_1)(\Gn+t_2)\Gr^+(\Gn)}=\fr{(\Gn-t_1)(\Gn-t_2)}{\Gr^-(\Gn)}[X^-(\Gn)]^{-1}\GF^-(\Gn) 
=R(\Gn), \quad \Gn\in {\Bbb C},
\label{3.66}
\eeq
where $\nu=-\Gd_*\csc^2\Gb$, 
and $R(\Gn)$ is a rational vector-function. Note that due to (\ref{3.60.5})
the vector $R(\Gn)$ has simple poles at the points $\pm\Gs_1$. Also, because of
 (\ref{3.60.10})
the vector $R(\Gn)$ has simple poles at the points $\Gn=\Gk_0\Gr_0$ in the case $\Gk_0=\pm 1$
and is bounded if $\Gk_0=0$. 
In addition, it has to have 
simple poles (the geometric optics poles) at the points $\Gn=\pm\Gn_0$. Since the vectors $\GF^\pm(\Gn)$ vanish,
the matrices $[X^\pm(\Gn)]^{-1}$ are bounded at infinity, and the elements of the matrix $G^1(\Gn)$ are degree-2 polynomials, the vector
$R(\Gn)$ has a pole at the infinite point of multiplicity 1 if $\Gk_0=0$ and multiplicity 2 if $\Gk_0=\pm 1$. 
The most general form of the vector $R(\Gn)$ in the former case is given by
\beq
R(\Gn)=\fr{P(\Gn)}{(\Gn^2-\Gs_1^2)(\Gn^2-\Gn_0^2)},
\label{3.66.1}
\eeq
where $P(\Gn)$ is an order-2 vector whose components, $P_1(\Gn)$ and $P_2(\Gn)$, are
degree-5 polynomials. In the case $\Gk_0=\pm 1$,
\beq
R(\Gn)=\fr{P(\Gn)}{(\Gn^2-\Gs_1^2)(\Gn^2-\Gn_0^2)(\Gn-\Gk_0\Gr_0)},
\label{3.66.1'}
\eeq
where the components $P_1(\Gn)$ and $P_2(\Gn)$ of the vector $P(\Gn)$ are degree-6 polynomials.
The vectors $\GF^\pm(\Gn)$ can be deduced from (\ref{3.66})
$$
\GF^+(\Gn)=\fr{\nu\Gr^+(\Gn)G^1(\Gn)X^+(\Gn)R(\Gn)}{(\Gn+\tau_1)(\Gn+\tau_2)(\Gn-t_1)(\Gn-t_2)},
\quad \Gn\in{\Bbb C}^+,
 $$
 \beq
 \GF^-(\Gn)=\fr{\Gr^-(\Gn)X^-(\Gn)R(\Gn)}{(\Gn-t_1)(\Gn-t_2)},\quad \Gn\in{\Bbb C}^-.
 \label{3.66.2}
 \eeq
The solution has twelve arbitrary constants, the coefficients of the degree-5 polynomials
$P_1(\Gn)$ and $P_2(\Gn)$, in the case $\Gk_0=0$ and fourteen constants if $\Gk_0=\pm 1$. 
 
In the case (ii), when one of the zeros of the polynomial $\Gd_0(\Gn)$, $\tau_1$, lies in the upper half-plane and the other, $\tau_2$, is in the lower half-plane,
 $\Gd_0(\Gn)=-(\Gn-\tau_1)(\Gn+\tau_2)\sin^2\Gb$, and
the boundary condition of the RHP gives
\beq
\fr{\nu (\Gn+\tau_2)[X^+(\Gn)]^{-1}\hat G^1(\Gn)\GF^+(\Gn)}{(\Gn+t_1)(\Gn+t_2)\Gr^+(\Gn)}=\fr{(\Gn-t_1)(\Gn-t_2)}{(\Gn-\tau_1)\Gr^-(\Gn)}[X^-(\Gn)]^{-1}\GF^-(\Gn) 
=R(\Gn), \quad \Gn\in {\Bbb C}.
\label{3.66.3}
\eeq
The rational vector $R(\Gn)$ is given by (\ref{3.66.1}), where $P_1(\Gn)$ and $P_2(\Gn)$
are polynomials of the fourth degree in the case $\Gn_0=0$ and by (\ref{3.66.1'}) with
the fifth degree polynomials in the case $\Gk_0=\pm 1$. The rearrangement of the factors of the polynomial $\Gd_0(\Gn)$ 
produces the new solution
$$
\GF^+(\Gn)=\fr{\nu\Gr^+(\Gn)G^1(\Gn)X^+(\Gn)R(\Gn)}{(\Gn+\tau_2)(\Gn-t_1)(\Gn-t_2)},
\quad \Gn\in{\Bbb C}^+,
 $$
 \beq
 \GF^-(\Gn)=\fr{(\Gn-\tau_1)\Gr^-(\Gn)X^-(\Gn)R(\Gn)}{(\Gn-t_1)(\Gn-t_2)},\quad \Gn\in{\Bbb C}^-,
 \label{3.66.4}
 \eeq
which has either ten, or twelve arbitrary constants depending whether $\Gk_0=0$, or $\Gk_0=\pm 1$.

In the third case, when both zeros of the polynomial $\Gd_0(\Gn)$ lie in the upper half-plane, 
$\Gd_0(\Gn)=-(\Gn-\tau_1)(\Gn-\tau_2)\sin^2\Gb$,
we have
\beq
\fr{\nu [X^+(\Gn)]^{-1}\hat G^1(\Gn)\GF^+(\Gn)}{(\Gn+t_1)(\Gn+t_2)\Gr^+(\Gn)}=\fr{(\Gn-t_1)(\Gn-t_2)}{(\Gn-\tau_1)(\Gn-\tau_2)\Gr^-(\Gn)}[X^-(\Gn)]^{-1}\GF^-(\Gn) 
=R(\Gn), \quad \Gn\in {\Bbb C}.
\label{3.66.5}
\eeq
As before, the rational vector $R(\Gn)$ has the form (\ref{3.66.1})
for $\Gk_0=0$ and (\ref{3.66.1'})
for $\Gk_0=\pm 1$. However, the components
of the vector $P(\Gn)$ are degree-3 polynomials if $\Gk_0=0$ and degree-4 polynomials if $\Gk_0=\pm 1$. The solution has eight arbitrary constants in the former case and ten
constants in the second case. It may  be written as
$$
\GF^+(\Gn)=\fr{\nu\Gr^+(\Gn)G^1(\Gn)X^+(\Gn)R(\Gn)}{(\Gn-t_1)(\Gn-t_2)},
\quad \Gn\in{\Bbb C}^+,
 $$
 \beq
 \GF^-(\Gn)=\fr{(\Gn-\tau_1)(\Gn-\tau_2)\Gr^-(\Gn)X^-(\Gn)R(\Gn)}{(\Gn-t_1)(\Gn-t_2)},\quad \Gn\in{\Bbb C}^-.
 \label{3.66.6}
 \eeq

\subsection{Symmetry conditions}
 
 In general, the solution derived in the previous section does not meet the symmetry condition $\GF^+(\Gn)=\GF^-(-\Gn)$, $\Gn\in{\Bbb C}$. We begin with the  case (i).
 In order to satisfy the symmetry condition,
  we rewrite it as
 \beq
 R(-\Gn)=\fr{\nu(\Gn+t_1)(\Gn+t_2)[\Gr^+(\Gn)]^2}{(\Gn+\tau_1)(\Gn+\tau_2)
 (\Gn-t_1)(\Gn-t_2)}
[X^-(-\Gn)]^{-1}G^1(\Gn)X^+(\Gn)R(\Gn),\quad \Gn\in{\Bbb C}^+.
\label{3.67}
\eeq 
Here, we used the relation $\Gr^+(\Gn)=[\Gr^-(-\Gn)]^{-1}$, $ \Gn\in{\Bbb C}^+$.
We wish now to simplify the matrix 
$U(\Gn)=[X^-(-\Gn)]^{-1}G^1(\Gn)X^+(\Gn)$.
It will be convenient to represent the function $\psi_1(\Gn)$ as the sum of the
even and odd functions
\beq
\psi_1(\Gn)=\psi_{1o}(\Gn)+\psi_{1e}(\Gn),\quad
\psi_{1o}(\Gn)=\fr{\Gk_0}{2}\log\fr{\Gn-\Gr_0}{\Gn+\Gr_0},
\quad
\psi_{1e}(\Gn)=\fr12\log\fr{\Gn^2-\Gs_1^2}{\Gn^2-\Gs_0^2},
\label{3.67'}
\eeq
(the function $\psi_{1o}(\Gn)\equiv 0$ if $\Gk_0=0$)
and use the following notations ($\psi_2(\Gn)$ is an even function):
$$
c=\cosh[f^{1/2}(\Gn)\psi_2(\Gn)], \quad  s=\sinh[f^{1/2}(\Gn)\psi_2(\Gn)], 
$$
\beq
Q(\pm\Gn)=
\left(\begin{array}{cc}
l_\pm& m_\pm\\
n_\pm & -l_\pm  \\
\end{array}
\right).
 \label{3.68}
 \eeq
By referring now to (\ref{3.60}), we transform the matrix $U(\Gn)$ as
\beq
U(\Gn)=e^{2\psi_{1o}(\Gn)}
 \left(\begin{array}{cc}
c-sl_-/\sqrt{f} \;    & -sm_-/\sqrt{f}  \\
 -sn_-/\sqrt{f}\;    & c+sl_-/\sqrt{f}       \\
\end{array}
\right)
  \left(\begin{array}{cc}
g_{11}^1   \;  &  g_{12}^1 \\
 g_{21}^1 \;  & g_{22}^1      \\
\end{array}
\right)
 \left(\begin{array}{cc}
c+sl_+/\sqrt{f}   \;  & sm_+/\sqrt{f}  \\
sn_+/\sqrt{f}  \;  & c-sl_+/\sqrt{f}     \\
\end{array}
\right).
 \label{3.69}
 \eeq
On using next the directly verified identities
$$
(l_+-l_-)g_{11}^1+n_+g_{12}^1-m_-g^1_{21}=0,\quad m_+g_{11}^1-(l_++l_-)g_{12}^1-m_-g_{22}^1=0,
 $$
$$ 
-n_-g_{11}^1+(l_++l_-)g_{21}^1+n_+g_{22}^1=0,\quad -n_-g_{12}^1+m_+g_{21}^1-(l_+-l_-)g_{22}^1=0,
$$
$$
c^2g_{11}^1+\fr{s^2}{f}(-l_-l_+g_{11}^1-l_-n_+g_{12}^1-l_+m_-g_{21}^1-m_-n_+g_{22}^1)=g_{11}^1,
$$$$
c^2g_{12}^1+\fr{s^2}{f}(-l_-m_+g_{11}^1+l_-l_+g_{12}^1-m_-m_+g_{21}^1+l_+m_-g_{22}^1)=g_{12}^1,
$$
$$
c^2g_{21}^1+\fr{s^2}{f}(-l_+n_-g_{11}^1-n_-n_+g_{12}^1+l_-l_+g_{21}^1+l_-n_+g_{22}^1)=g_{21}^1,
$$
\beq
c^2g_{22}^1+\fr{s^2}{f}(-m_+n_-g_{11}^1+l_+n_-g_{12}^1+l_-m_+g_{21}^1-l_-l_+g_{22}^1)=g_{22}^1,
\label{3.70}
\eeq
we ultimately deduce that $\exp\{-2\psi_{1o}(\Gn)\}U(\Gn)$ is a polynomial matrix that
 coincides with $G^1(\Gn)$. Furthermore,
\beq
[X^-(-\Gn)]^{-1}G^1(\Gn)X^+(\Gn)=\left(\fr{\Gn-\Gr_0}{\Gn+\Gr_0}\right)^{\Gk_0}G^1(\Gn),\quad \Gn\in{\Bbb C^+}.
\label{3.71}
 \eeq
Next, for the present purpose, we  transform the symmetry condition (\ref{3.67}) as
\beq
R(-\Gn)=\fr{\nu }{(\Gn-t_1)(\Gn-t_2)}\left(\fr{\Gn-\Gr_0}{\Gn+\Gr_0}\right)^{\Gk_0}G^1(\Gn)R(\Gn)
\label{3.79} 
 \eeq
or, equivalently,
\beq
P(-\Gn)=\fr{(-1)^{\Gk_0}\nu G^1(\Gn)P(\Gn)}{(\Gn-t_1)(\Gn-t_2)},
\label{3.80} 
 \eeq
where the components of the vector $P(\Gn)$, $P_1(\Gn)$ and $P_2(\Gn)$, are polynomials
\beq
P_1(\Gn)=\sum_{j=0}^{|\Gk_0|+5}a_j\Gn^j, \quad P_2(\Gn)=\sum_{j=0}^{|\Gk_0|+5}b_j\Gn^j
\label{3.81}
\eeq
whose coefficients are to be determined. 
On replacing $\Gn$ by $-\Gn$ we reduce the  condition (\ref{3.80}) to 
\beq
P(-\Gn)=\fr{(\Gn+t_1)(\Gn+t_2)}{(-1)^{\Gk_0}\nu}[G^1(-\Gn)]^{-1}P(\Gn)
\label{3.83}
\eeq
which is equivalent to the condition (\ref{3.80}) since the matrix $G^1(\Gn)$ possesses the following property:
\beq
G^1(\Gn)G^1(-\Gn)=d_1(\Gn)I.
\label{3.84}
\eeq
The symmetry condition (\ref{3.80}) asserts that the coefficients of the polynomials $P_1(\Gn)$ and $P_2(\Gn)$
can be determined from the two equations
\beq
Q_1(\Gn)=0, \quad Q_2(\Gn)=0,
\label{3.85}
\eeq
where
\beq
Q_j(\Gn)=\sin^2\Gb(\Gn-t_1)(\Gn-t_2)P_j(-\Gn)+(-1)^{\Gk_0}\Gd_*
[g_{j1}^1(\Gn)P_1(\Gn)+g_{j2}^1(\Gn)P_2(\Gn)], \quad j=1,2,
\label{3.86}
\eeq
are polynomials of degree $6+|\Gk_0|$ (the terms $\Gn^{7+|\Gk_0|}$ in both polynomials have zero-coefficients). 

Consider first the case $\Gk_0=0$.
To define the coefficients, we compute the derivatives
$Q_j^{(m)}(\Gn)$, $m=0,1,\ldots,6; j=1,2$. The first six equations
$Q_1^{(6-m)}(\Gn)=0$ ($m=0,1,\ldots,5$) express the coefficients $b_{5-m}$ through the coefficients of the polynomial $P_1(\Gn)$
$$
b_{2j}=\nu_0(-a_{2j}\nu_1^- + a_{2j+1} \nu_2^+),
$$
\beq
b_{2j+1}=\nu_0(-2 a_{2j} \sin^2\Gb - a_{2j+1}\nu_1^++a_{2j+2}\nu_2^-), \quad j=0,1,2, 
\label{3.87}
\eeq
where
$$
\nu_0=\fr{\Gg_1^+ + \Gg_4^+}{2\cos\Gb[(\Gg^+_4)^2-\Gg_1^-\Gg_4^-+k^2\sin^2\Gb]}, \quad \nu_1^\pm=\Gg_1^- - \Gg_4^-\pm(t_1 + t_2) \sin^2\Gb,  
$$
\beq
\nu_2^\pm=\Gg_1^- \Gg_4^- + k^2 \cos^2\Gb \pm t_1 t_2 \sin^2\Gb, \quad a_6=0.
\label{3.88}
\eeq
The equation $Q_1(\Gn)=0$ yields
\beq
a_0[2 \Gg_1^-\Gg_4^- + k^2\cos^2\Gb - t_1 t_2\sin^2\Gb]=0.
\label{3.89}
\eeq
It is directly verified that $2 \Gg_1^-\Gg_4^- + k^2\cos^2\Gb - t_1 t_2\sin^2\Gb=0$ and therefore $a_0$ is free. The other seven equations $Q_2^{(j)}=0$ ($j=0,1,\ldots,6$) are also identically
satisfied provided the coefficients $b_j$ ($j=0,1,\ldots,5$) are chosen as in (\ref{3.87}),
and $t_1$, $t_2$ are the two zeros in the upper-half plane of the even degree-4 polynomial $d_1(\Gn)$. 

A similar result holds for the case $|\Gk_0|=1$. The equations $Q_1^{(7-m)}(\Gn)=0$ ($m=0,1,\ldots,6$)   express the coefficients
$b_{6-m}$ through the coefficients of the polynomial $P_1(\Gn)$
$$
b_{2j}=\nu_0(-2 a_{2j-1} \sin^2\Gb - a_{2j}\nu_1^++a_{2j+1}\nu_2^-), \quad j=0,1,2,3, \quad a_{-1}=a_7=0,
$$
\beq
b_{2j+1}=\nu_0(-a_{2j+1}\nu_1^- + a_{2j+2} \nu_2^+),\quad j=0,1,2.
\label{3.89.1}
\eeq
In the case $\Gk_0=\pm 1$, we have
$2 \Gg_1^-\Gg_4^- + k^2\cos^2\Gb + t_1 t_2\sin^2\Gb=0$. Therefore,  the equation  $Q_1(\Gn)=0$, which is equivalent to
\beq
 a_0[2 \Gg_1^-\Gg_4^- + k^2\cos^2\Gb +t_1 t_2\sin^2\Gb]=0,
 \label{3.89.2}
\eeq
is satisfied for any $a_0$. If the coefficients  $b_j$ are defined by (\ref{3.89.1}), then $Q_2(\Gn)\equiv 0$. 
Consequently, in the case (i), the solution to the vector RHP 
possesses $6+|\Gk_0|$ arbitrary constants  $a_j$ ($j=0,1,\ldots,5+|\Gk_0|$) and satisfies the symmetry condition $\GF^+(\Gn)=\GF^-(-\Gn)$.

Consider next the case (ii) when the two zeros of the polynomial $\Gd_0(\Gn)$
lie in the opposite half-planes.  Since the relation (\ref{3.71})
is invariant with respect to the location of the zeros of the polynomial $\Gd_0(\Gn)$,
we employ it again and deduce from (\ref{3.66.4}) 
\beq
R(-\Gn)=-\fr{\nu }{(\Gn-t_1)(\Gn-t_2)}\left(\fr{\Gn-\Gr_0}{\Gn+\Gr_0}\right)^{\Gk_0}G^1(\Gn)R(\Gn).
\label{3.89.3} 
 \eeq
The rational vector $R(\Gn)$ is given by (\ref{3.66.1}) if $\Gk_0=0$ and by (\ref{3.66.1'})
if $\Gk_0=\pm 1$, and
\beq
P_1(\Gn)=\sum_{j=0}^{|\Gk_0|+4}a_j\Gn^j, \quad P_2(\Gn)=\sum_{j=0}^{|\Gk_0|+4}b_j\Gn^j.
\label{3.89.4}
\eeq
The symmetry condition  (\ref{3.89.3}) reads for the polynomials $P_1(\Gn)$ and $P_2(\Gn)$
\beq
P(-\Gn)=\fr{(-1)^{\Gk_0+1}\nu G^1(\Gn)P(\Gn)}{(\Gn-t_1)(\Gn-t_2)}.
\label{3.89.4'}
\eeq
As in our earlier derivations  described for the case (i), we obtain for $\Gk_0=0$
$$
b_{2j}=\nu_0(-2 a_{2j-1} \sin^2\Gb - a_{2j}\nu_1^++a_{2j+1}\nu_2^-), \quad j=0,1,2, \quad a_{-1}=a_5=0,
$$
\beq
b_{2j+1}=\nu_0(-a_{2j+1}\nu_1^- + a_{2j+2} \nu_2^+),\quad j=0,1.
\label{3.89.5}
\eeq
If $|\Gk_0|=1$, then the coefficients $b_{2j}$ and $b_{2j+1}$ coincide with the corresponding coefficients in the case (i), $\Gk=0$, and
are defined  by (\ref{3.87}).
In the case (ii), the solution is given by (\ref{3.66.4}), 
(\ref{3.66.1}), (\ref{3.66.1'}), (\ref{3.89.4}), (\ref{3.89.5}) and (\ref{3.87}). It satisfies the symmetry
conditions and possesses $5+|\Gk_0|$ arbitrary constants $a_0, a_1,\ldots, a_{4+|\Gk_0|}$. 

Finally, in the case (iii), when the two zeros of the polynomial $\Gd_0(\Gn)$
lie in the upper half-plane, the symmetry relation coincides with the one derived in the case (i)
and is given by 
(\ref{3.79}). The polynomials $P_1(\Gn)$ and $P_2(\Gn)$ in the representation 
of the solution (\ref{3.66.6}) have the form
\beq
P_1(\Gn)=\sum_{j=0}^{|\Gk_0|+3}a_j\Gn^j, \quad P_2(\Gn)=\sum_{j=0}^{|\Gk_0|+3}b_j\Gn^j,
\label{3.89.7}
\eeq
where the constants $a_j$ ($j=0,1,\ldots,3+|\Gk_0|$) are free and if $\Gk_0=0$,
$$
b_{2j}=\nu_0(-a_{2j}\nu_1^- + a_{2j+1} \nu_2^+),
$$
\beq
b_{2j+1}=\nu_0(-2 a_{2j} \sin^2\Gb - a_{2j+1}\nu_1^++a_{2j+2}\nu_2^-), \quad j=0,1, \quad  a_4=0.
\label{3.89.8}
\eeq
If $|\Gk_0|=1$, the coefficients are defined by (\ref{3.89.5}).
The solution derived has $4+|\Gk_0|$ arbitrary constants
 $a_0, a_1,\ldots, a_{3+|\Gk_0|}$.

\subsection{Additional mathematical conditions}

Since the vector polynomial $P(\Gn)$ meets the condition (\ref{3.80}) in the  cases (i) and (iii) and the condition (\ref{3.89.4'}) in the case (ii) for all $\Gn$, and the left-hand side in (\ref{3.80}) and  (\ref{3.89.4'}) 
is bounded at the points $t_1$ and $t_2$,  due to (\ref{3.66.2}),  (\ref{3.66.4}) and  (\ref{3.66.6}),
the vector $\GF^+(\Gn)$, regardless of which case is considered, (i), (ii), or (iii),  has removable singularities at the points $t_1$ and $t_2$
($\I t_j>0$). 

Now, the property (\ref{3.60.5}) of the matrix $[X(\Gn)]^{-1}$ asserts that at the point
$\Gs_1$
the polynomials $P_1(\Gn)$ and $P_2(\Gn)$ satisfy the relation
\beq
s^+_j P_1(\Gs_1)=P_2(\Gs_1),
\label{3.91} 
\eeq
where $s_j^+=n(\Gs_1)[l(\Gs_1)+(-1)^{j-1}\sqrt{f(\Gs_1)}]^{-1}$,
$j=1$ if $p_1\in{\Bbb C}_1$, and  $j=2$ if $p_1\in{\Bbb C}_2$. This relation is necessary and sufficient for the removal of the simple pole of the solution
at the point $\Gs_1$.
Because of the symmetry condition $\GF^+(\Gn)=\GF^-(-\Gn)$, $\Gn\in{\Bbb C}^+$,
equation (\ref{3.91}) removes the singularity of the vector $\GF^-(\Gn)$
at the point $-\Gs_1$ as well.

According to the property (\ref{3.60.2}) the solution to the vector RHP
has simple poles at the points $\pm\Gs_0$. We may select $\Gs_0$, the affix of the arbitrary fixed point
$p_0$ on the first sheet of the surface $\frak R$, such that $\I\Gs_0>0$.
Since $Y^+$  in (\ref{3.60.2}) is a rank-1 matrix,
the point $\Gs_0$ is a removable point of the vector $\GF^+(\Gn)$ if and only if the
following condition holds:
\beq
Y^+_{11}P_1(\Gs_0)+Y^+_{12}P_2(\Gs_0)=0,
\label{3.92} 
\eeq
where $(Y_{11}^+,Y_{12}^+)$ is the first row of the matrix 
$Y(\Gs_0,\sqrt{f(\Gs_0)})$. Again, due to the symmetry, this condition implies that
the point $-\Gs_0$ is also a removable point of the solution.

When $\Gk_0=1$,  according
to (\ref{3.60.10}) and (\ref{3.66.1'}), the vector $\GF^+(\Gn)$ has inadmissible pole at the point 
$\Gn=\Gr_0\in{\Bbb C^+}$ (the point $\Gr_0$ was chosen to be in the upper half-plane) due to the pole of the vector $R(\Gn)$. 
The vector $\GF^-(\Gn)$ on the other hand, due to (\ref{3.60.10}), has a simple pole at the point $\Gn=-\Gr_0$.
Since $\tilde Y^-$ in (\ref{3.60.10}) is a  rank-1 matrix and the solution is symmetric,
for the points
$\Gn=\pm\Gr_0\in{\Bbb C^\pm}$ being removable points of $\GF^\pm(\Gn)$ it is necessary and sufficient that
the following condition holds:
\beq
\tilde Y^-_{11}P_1(-\Gr_0)+\tilde Y^-_{12}P_2(-\Gr_0)=0,
\label{3.95}
\eeq
where $\{\tilde Y^-_{11},\tilde Y^-_{12}\}$ is the first row of the matrix
$Y(-\Gr_0,-\sqrt{f(\Gr_0)})$.

In the case $\Gk_0=-1$, the matrix $X(\Gn)$ has a simple pole 
at the point $\Gn=\Gr_0$, and the vector $R(\Gn)$ has a simple pole at $\Gn=-\Gr_0$. Therefore, the solution
$\GF^\pm(\Gn)$ has simple poles at the points $\pm\Gr_0$.
As in the previous case, they can be removed by a single condition. It has the form
\beq
\tilde Y^+_{11}P_1(\Gr_0)+\tilde Y^+_{12}P_2(\Gr_0)=0,
\label{3.95'}
\eeq
where $\{\tilde Y^+_{11},\tilde Y^+_{12}\}$ is the first row of the matrix
$Y(\Gr_0,\sqrt{f(\Gr_0)})$.
The solution derived in the previous section has $s+|\Gk_0|$
arbitrary constants. On the other hand, if $\Gk_0=0$, the points $\pm\Gr_0$ are regular points, and if  $\Gk_0=\pm 1$, they are poles removed by one 
condition. This asserts that the integer $\Gk_0$ does not effect the number 
of solutions of the vector RHP. 
This completes  the solution procedure for the vector RHP. 

Since the structure 
of the matrix coefficient $\hat G(\Gn)$ for the RHP 2 is the same as for the RHP 1
(the five parameters 
of the problem, $\Gg_1^\pm$, $\Gg_4^\pm$ and $\Gb$,  
need to be changed according to the transformation (\ref{2.18})), the same results are valid for
the RHP 2.  Recall that $\det G(\Gn)=\Gd_0(-\Gn)/\Gd_0(\Gn)$.
Then $\det \hat G(\Gn)=\hat\Gd_0(-\Gn)/\hat\Gd_0(\Gn)$, where
\beq
\hat\Gd_0(\Gn)=(\Gn^2-k_0^2)\cos^2\Gb-(\Gn-\Gg_1^+)(\Gn-\Gg_4^+).
\label{3.96.1}
\eeq
Denote the zeros of the polynomials $\Gd(\Gn)$ and $\hat\Gd(\Gn)$ as $\Gn_j$ and $\hat\Gn_j$ ($j=1,2$), respectively, 
\beq
\Gn_j=\fr{\Gg_1^-+\Gg_4^-+(-1)^{j-1}\sqrt{\GD_-}}{2\sin^2\Gb},\quad 
\hat\Gn_j=\fr{\Gg_1^++\Gg_4^++(-1)^{j-1}\sqrt{\GD_+}}{2\sin^2\Gb},\quad j=1,2,
\label{6.37'}
\eeq
where
\beq
\GD_\pm=(\Gg_1^\pm-\Gg_4^\pm)^2+4(\Gg_1^\pm\Gg_4^\pm-k_0^2\sin^2\Gb)\cos^2\Gb.
\label{6.38}
\eeq

\vspace{.1in}

Now we  summarize the results.

\vspace{.1in}

{\sl Theorem 5.1. Let 
\beq
2\Gk=\fr{1}{2\pi}[\arg\det G(\Gn)]|_{-\infty}^{+\infty}, \quad 2\hat\Gk=\fr{1}{2\pi}[\arg\det \hat G(\Gn)]|_{-\infty}^{+\infty}
\label{3.95.0}
\eeq
and
\beq
\Gk_j=\left\{
\begin{array}{cc}
1, & \I\Gn_j<0, \\
0,  & \I\Gn_j>0,\\
\end{array}
\right.
\quad \hat\Gk_j=\left\{
\begin{array}{cc}
1, & \I\hat\Gn_j<0, \\
0,  & \I\hat\Gn_j>0.\\
\end{array}
\right.
\label{3.95.1}
\eeq

Then

(i) $\Gk=\Gk_1+\Gk_2-1$, $\hat\Gk=\hat\Gk_1+\hat\Gk_2-1$, 

(ii) the  solution to the vector RHPs 1 and 2  exists and possesses $\Gk+3$  and $\hat\Gk+3$ arbitrary constants, respectively.
}

\vspace{.1in}

Now, the solutions to the vector RHPs 1 and 2 need to be compatible. Indeed, from equations (\ref{3.2}) ($\Gg_2^+=\Gg_3^+=0$)
 we can express $\hat\Gf_j(i\Gz,0)$ through the solution to the vector RHP 1 
  \beq
\hat\Gf_j(i\Gz,0)=-\fr{\Gg_{j+}+i\Gz}{2\Gn}[\GF_j^+(\Gn)-\GF_j^-(\Gn)]-\fr{(-1)^j\cos\Gb}{2}[\GF_{3-j}^+(\Gn)+\GF_{3-j}^-(\Gn)], \quad j=1,2,
\label{3.95.2}
\eeq
where $\Gg_{1+}=\Gg_1^+$, $\Gg_{2+}=\Gg_4^+$.
On replacing $i\Gz$ by $\Gn$  we obtain 
 the functions $\hat\GF_j^\pm(\Gn)$. They constitute  the solution of the vector RHP 2 but have a  set of $\Gk+3$ free constants of the RHP 1. 
 On the other hand, the functions
$\hat\GF_j^\pm(\Gn)$ can be derived directly by solving the RHP 2. These new expressions will have their own set of  $\hat\Gk+3$ free constants.
The two solutions have to be the same, and we require
$$
\left.
-\fr{\Gg_{j+}+\Gn}{2\hat\Gz}[\GF_j^+(\hat\Gz)-\GF_j^-(\hat\Gz)]-\fr{(-1)^j\cos\Gb}{2}[\GF_{3-j}^+(\hat\Gz)+\GF_{3-j}^-(\hat\Gz)]\right|_{RHP\, 1}
$$
\beq
\left.=\hat\GF_j^+(\Gn)\right|_{RHP \,2}, \quad j=1,2.
\label{3.95.3}
\eeq
Here, $\hat\Gz$ is a  branch of the two-valued function $\hat\Gz^2=k_0^2-\Gn^2$. We   fix it 
by the condition $\hat\Gz(0)=k_0$ in the
$\Gn$-plane cut along the lines passing through the infinite points and joining the branch points $\pm k_0$ 
(formulas  (\ref{3.95.3}) are invariant with respect to the choice of the branch $\hat\Gz$).
We conjecture that  the compatibility conditions (\ref{3.95.3})  comprise $\Gk+\hat\Gk+4$ equations for $\Gk+\hat\Gk+6$ constants.
Furthermore, in the case of normal incidence,

(i) if $\Gk_1=\Gk_2=\hat\Gk_1=\hat\Gk_2=1$, then each equation in (\ref{3.95.3}) yields  three conditions
which eliminate six constants,

(ii) if $\Gk_1=\Gk_2=\hat\Gk_1=1$, $\hat\Gk_2=0$, then the first and second equations  give three and two  conditions, respectively,

(iii) if $\Gk_1=\Gk_2=1$,  $\hat\Gk_1=\hat\Gk_2=0$, then  each  equation in  (\ref{3.95.3}) eliminate two
constants, 

(iv) if $\Gk_1=1$, $\Gk_2=0$, $\hat\Gk_1=1$, $\hat\Gk_2=0$, then the first equation  gives three  conditions for the unknown constants, while the second one gives only one,

(v) if $\Gk_1=1$, $\Gk_2=\hat\Gk_1=\hat\Gk_2=0$, then the first and  second equations  give two
and one  conditions, respectively,

(vi) if $\Gk_1=\Gk_2=\hat\Gk_1=\hat\Gk_2=0$, then there are two arbitrary constants say, $D_j$ (RHP 1)
and $\hat D_j$ (RHP 2), in the solution to each RHP, and  $D_j=c_j \hat D_j,$ $j=1,2$.
 Each equation in (\ref{3.95.3}) brings us just one equation defining the scale factor $c_j$.
 Therefore, the compatibility conditions can be discarded.

We will verify this statement in section 7 by using the method of undetermined 
coefficients. It is possible that the same statement is true in the general case.
To apply this   method  for the oblique incidence, it it is required to implement some tedious computations.  
For numerical purposes, it is sufficient to solve equations  (\ref{3.95.3}) at arbitrary distinct fixed points. The number of points and equations
is equal to $\Gk+\hat\Gk+4$, and the equations need to be chosen such that
the determinant of the system does not vanish.

We also note that the vectors $\GF^+(\Gn_0)$ and $\hat\GF^+(\Gn)$ found have simple poles 
at the geometric optics poles $\pm\Gn_0$ and $\pm\hat\Gn_0$, respectively.
The residues of the solution at these
points have not been specified. In the next section we fix the residues              
and recover the incident and reflected waves. Since the compatibility conditions have been satisfied,
both RHPs have two free constants, and it is sufficient to work with the solution to the RHP 1 only.

\setcounter{equation}{0}

\section{Sommerfeld integral representation: solution to the diffraction problem}

The electric and magnetic field can be given in terms of the Sommerfeld integrals as
 \beq
\left(\begin{array}{c}
 E_z (\Gr,\Gt)\\
 ZH_z(\Gr,\Gt)\\
 \end{array}
 \right)=\fr{1}{2\pi i}\int_\CT e^{-i k_0\Gr\cos s}
 \left(\begin{array}{c}
S_1(s+\Gt)\\
S_2(s+\Gt)\\
 \end{array}
 \right)ds,
 \label{6.1}
 \eeq
where $\CT$ is the Sommerfeld double loop with the asymptotes $\R s=\pi/2$ and $\R s=-3\pi/2$ for the upper
loop $\CT_+$ lying in the upper half-plane. The starting point  of the contour is chosen to be    
$s=\pi/2+i\infty$. The second loop $\CT_-$ is symmetric to $\CT_+$ with respect to the origin.

\subsection{Spectra $S_1(s)$ and $S_2(s)$}

 The purpose of this section is to derive expressions for the functions $S_1(s)$ and $S_2(s)$. 
 They are analytic everywhere in the strip $|\R s|<\pi/2+\Ge$ ($\Ge>0$ and small) except for
 the simple poles $s=\pm\Gt_0$, and
 \beq
  \mathop{\rm res}\limits_{s=\Gt_0}S_1(s)=i_1,\quad    \mathop{\rm res}\limits_{s=\Gt_0}S_2(s)=i_2.
\label{6.1'}
\eeq 
 The knowledge  
of these functions enables us to fix the residues of the solution to the RHPs 1 and 2 at the geometric
optics poles and recover the incident, reflected, surface and diffracted fields. 
On applying the inverse Maliuzhinets transform {\bf(\ref{mal2})},
\beq
S_j(\Gt+s)-S_j(\Gt-s)=ik_0\sin s\int_0^\infty e^{ik_0\Gr\cos s}\Gf_j(\Gr,\Gt)d\Gr,
\label{6.2}
\eeq
and fixing first $\Gt=0$ and then $\Gt=\pi/2$, we  obtain the following two equations:
$$
S_j(s)-S_j(-s)=F_{j-}(s),
$$
\beq
S_j\left(\fr{\pi}{2}+s\right)-S_j\left(\fr{\pi}{2}-s\right)=F_{j+}(s),\quad j=1,2,
\label{6.3}
\eeq
where
\beq
F_{j-}(s)=ik_0\sin s\int_0^\infty e^{ik_0 x\cos s}\Gf_j(x,0)dx,\quad
F_{j+}(s)=ik_0\sin s\int_0^\infty e^{ik_0 y\cos s}\Gf_j(0,y)dy.
\label{6.3'}
\eeq
We will adopt the Fourier transform in the form {\bf(\ref{mal2})}  and use the following notations for the transforms and their inverses:
$$
\hat S_j(t)=-\fr{1}{2\pi}\int_{-i\infty}^{i\infty} S_j(s) e^{ist}ds, \quad S_j(s)=\int_{-i\infty}^{i\infty} \hat S_j(t) e^{-ist}dt, 
$$
\beq 
\hat F_{j\pm}(t)=-\fr{1}{2\pi}\int_{-i\infty}^{i\infty} F_{j\pm}(s) e^{ist}ds, \quad F_{j\pm}(s)=\int_{-i\infty}^{i\infty} \hat F_{j\pm}(t) e^{-ist}dt. 
 \label{6.6}
 \eeq
When applied to the functions $S_j(\pi/2\pm s)$, this transformation yields
$$
-\fr{1}{2\pi}\int_{-i\infty}^{i\infty} S_j\left(\fr{\pi}{2}\pm s\right) e^{ist}ds=-\fr{e^{\mp i\pi t/2}}{2\pi}\int_{\pi/2-i\infty}^{\pi/2+i\infty}
 S_{j}(s) e^{\pm ist}ds
 $$
 \beq
 =\hat S_j(\pm t)e^{\mp i\pi t/2}-ii_j e^{\pm i(\Gt_0-\pi/2)t}.
 \label{6.7} 
\eeq
Here we used the fact that the functions $S_j(s)$ are analytic everywhere in the strip $|\R s|<\pi/2+\Ge$ apart from a simple pole 
at the point $s=\Gt_0$ with the residue  given by (\ref{6.1'}).
 From these we find that corresponding to (\ref{6.3}) we have the pair of equations
 $$
 \hat S_j(t)-\hat S_j(-t)=\hat F_{j-}(t),
$$
\beq
 \hat S_j(t)e^{-i\pi t/2}-\hat S_j(-t)e^{i\pi t/2}=\hat F_{j+}(t)+2i_j\sin\left(\fr{\pi}{2}-\Gt_0\right)t.
 \label{6.8} 
 \eeq
Because of the symmetry of these equations it suffices to determine one of the functions $\hat S_j(\pm t)$
say, $\hat S_j(t)$, 
\beq
\hat S_j(t)=\fr{\hat F_{j-}(t)e^{i\pi t/2}-\hat F_{j+}(t)}{2i\sin\pi t/2}+\fr{ii_j\sin(\pi/2-\Gt_0)t}{\sin\pi t/2}.
\label{6.9}
\eeq
It only remains to apply the inverse transform (\ref{6.6}) and use the integral
\beq
\int_{-i\infty}^{i\infty}\fr{e^{i\Gs t}dt}{\sin\pi t/2}=-2\tan\Gs, \quad -\fr{\pi}{2}<\R \Gs<\fr{\pi}{2},
\label{6.10}
\eeq
to obtain an integral representation for the function $S_j(s)$.  We can fashion a similar formula for the function associated with the magnetic field.
When combined the expressions become
\beq
S_j(s)=S_j^{(1)}(s)+S_j^{(2)}(s)+i_j[\cot(s-\Gt_0)-\cot(s+\Gt_0)], \quad 0<\R s<\fr{\pi}{2},
\label{6.11}
\eeq
where
$$
S_j^{(1)}(s)=-\fr{1}{2\pi i}\int_{-i\infty}^{i\infty}F_{j-}(\Gs)\cot(\Gs-s)d\Gs, \quad 0<\R s<\pi,
$$
\beq
S_j^{(2)}(s)=-\fr{1}{2\pi i}\int_{-i\infty}^{i\infty}F_{j+}(\Gs)\tan(\Gs-s)d\Gs, \quad -\fr{\pi}{2}<\R s<\fr{\pi}{2},
\label{6.12}
\eeq
The functions $S_1(s)$ and $S_2(s)$ are analytic everywhere in the strip $0<\R s<\pi/2$ except for the point $s=\Gt_0$, where
they have a simple pole with the residue equal to $i_1$ and $i_2$, respectively. 
The integrands in (\ref{6.12}) can be expressed through the solution to the vector RHP on employing  (\ref{6.3'})  and (\ref{3.2})
$$
F_{1-}(s)=-\fr{i}{2}(\Gg_{1}^++k_0\cos s)[\GF_1^+(k_0\sin s)-\GF_1^-(k_0\sin s)]
$$
$$
+\fr{i}{2}k_0\cos\Gb \sin s[\GF_2^+(k_0\sin s) +\GF_2^-(k_0\sin s)],
$$
$$
F_{2-}(s)=-\fr{i}{2}(\Gg_{4}^++k_0\cos s)[\GF_2^+(k_0\sin s)-\GF_2^-(k_0\sin s)]
$$
$$
-\fr{i}{2}k_0\cos\Gb \sin s[\GF_1^+(k_0\sin s) +\GF_1^-(k_0\sin s)],
$$
\beq
F_{j+}(s)=ik_0\sin s\GF_j^+(k_0\cos s), \quad j=1,2.
\label{6.5}
\eeq
Formula (\ref{6.11}) is remarkable for
the functions $S_j(s)$ being determined in the strip $0<\R s<\pi/2$ independently
of the residues of the solution of the RHP at the geometric optics poles.
In the next section we will show that in order to recover the reflected waves,  these residues need to be specified.

\subsection{Analytic continuation of the spectra}

Let
\beq
\GF_j^+(\Gn)\sim\fr{iC_j}{\Gn-\Gn_0}, \quad \Gn\to \Gn_0,
\label{6.13}
\eeq
where $C_j$  are some complex constants to be determined.
Since the functions $\GF_j^\pm(\Gn)$  have simple poles 
at the points $\pm\Gn_0$, the functions
$\GF_j^\pm(k_0\cos s)$ have simple poles at the points
$s=\pm(\pi/2-\Gt_0)+m\pi$, $s=\pm(\pi/2+\Gt_0)+m\pi$, $m\in{\Bbb Z}$, while the functions
$\GF_j^\pm(k_0\sin s)$  have simple poles $s=\pm\Gt_0+m\pi$,
$m\in{\Bbb Z}$. At the poles of interest, because of the symmetry conditions 
$\GF_j^+(\Gn)=\GF_j^-(-\Gn)$, we have
$$
 \mathop{\rm res}\limits_{s=\pm(\pi/2-\Gt_0)}\GF_j^+(k_0\cos s)=-\mathop{\rm res}\limits_{s=\pm(\pi/2
 +\Gt_0)}\GF_j^-(k_0\cos s)=
 \mp\fr{iC_j}{\hat\Gn_0},
 $$
 \beq
 \pm\mathop{\rm res}\limits_{s=\pm\Gt_0}\GF_j^\pm(k_0\sin s)=-\mathop{\rm res}\limits_{s=\pm\pi-
 \Gt_0}\GF_j^+(k_0\sin s)=\mathop{\rm res}\limits_{s=\pm\pi+
 \Gt_0}\GF_j^-(k_0\sin s)=\fr{iC_j}{\hat\Gn_0},\quad j=1,2.
\label{6.14}
\eeq
To determine the second  group  of residues associated with the reflected waves,
in addition to the symmetry conditions we employ the  boundary condition of the RHP to continue analytically the function
$\GF^-(\Gn)$  into the upper half-plane
\beq
\GF^-(\Gn)=[G(\Gn)]^{-1}\GF^+(\Gn),\quad 
 \Gn\in{\Bbb C^+}.
\label{6.15} 
\eeq
 This enables us to write
 $$
 \mathop{\rm res}\limits_{s=\pm(\pi/2+\Gt_0)}\GF^+(k_0\cos s)=-\mathop{\rm res}\limits_{s=\pm(\pi/2-\Gt_0)}\GF^-(k_0\cos s)
 =\pm\fr{i}{\hat\Gn_0}[G(\Gn_0)]^{-1}C,
$$ 
 \beq
\pm \mathop{\rm res}\limits_{s=\mp\Gt_0}\GF^\pm(k_0\sin s)=-\mathop{\rm res}\limits_{s=\pm\pi+\Gt_0}\GF^+(k_0\sin s)
 =\mathop{\rm res}\limits_{s=\pm\pi-\Gt_0}\GF^-(k_0\sin s)
 =-\fr{i}{\hat\Gn_0}[G(\Gn_0)]^{-1}C,
 \label{6.16}
 \eeq
 where $C=(C_1,C_2)^T$.

Next we continue 
analytically
the functions $S_j^{(1)}(s)$ from the strip $0<\R s<\pi$ to the strips $\pi<\R s < 3\pi/2$ and $-\pi<\R s<0$
and the functions $S_j^{(2)}(s)$ from the strip $-\pi/2<\R s<\pi/2$ to the strips $\pi/2<\R s < 3\pi/2$ and $-\pi<\R s<-\pi/2$. This procedure requires computing the residues of the functions $F_{j-}(s)$
 at the poles $\pm\Gt_0$ and $\pm(\pi-\Gt_0)$ and the functions  $F_{j+}(s)$ at the poles
 $\pm(\pi/2-\Gt_0)$ and $\pm(\pi/2+\Gt_0)$. Denote
 \beq
 [G(\Gn_0)]^{-1}=\left(\begin{array}{cc}
 \mu_{11} \; & \mu_{12}\\
 \mu_{21} \; & \mu_{22}\\
 \end{array}
 \right).
 \label{6.17} 
 \eeq
 On employing (\ref{6.14}) and (\ref{6.16}) we derive 
 $$
 \mathop{\rm res}\limits_{s=\pm\Gt_0} F_{j-}(s)=\GL_{j-}, \quad 
 \mathop{\rm res}\limits_{s=\pm(\pi-\Gt_0)} F_{j-}(s)=M_{j-},
 $$
 \beq
\mathop{\rm res}\limits_{s=\pm(\pi/2-\Gt_0)} F_{j+}(s)=\GL_{j+}, \quad 
 \mathop{\rm res}\limits_{s=\pm(\pi/2+\Gt_0)} F_{j+}(s)=M_{j+}, 
 \label{6.18}
 \eeq
 where
 $$
 \GL_{1+}=C_1,\quad  \GL_{2+}=C_1,\quad  \GL_{1-}=p_1^+C_1-q_1^+C_2,\quad  \GL_{2-}=-q_2^+C_1+p_2^+C_2,
 $$
 $$
 M_{1+}=-\mu_{11}C_1-\mu_{12}C_2, \quad  M_{2+}=-\mu_{21}C_1-\mu_{22}C_2,  
 $$$$
   M_{1-}=-p_{1}^-C_1+q_1^-C_2, 
 \quad  M_{2-}=q_{2}^-C_1-p_{2}^-C_2.  
  $$$$
 p_1^\pm=\fr{1}{2\hat\Gn_0}[(\Gg_1^+\pm\hat\Gn_0)(1-\mu_{11})-\Gn_0\mu_{21}\cos\Gb], \quad
 q_1^\pm=\fr{1}{2\hat\Gn_0}[(\Gg_1^+\pm\hat\Gn_0)\mu_{12}+\Gn_0(1+\mu_{22})\cos\Gb],
 $$
 \beq
 p_2^\pm=\fr{1}{2\hat\Gn_0}[(\Gg_4^+\pm\hat\Gn_0)(1-\mu_{22})+\Gn_0\mu_{12}\cos\Gb],\quad
 q_2^\pm=
 \fr{1}{2\hat\Gn_0}[(\Gg_4^+\pm\hat\Gn_0)\mu_{21}-\Gn_0(1+\mu_{11})\cos\Gb].
 \label{6.19}
\eeq
In addition, the functions $F_{j\pm}(s)$  may have some poles which generate surface waves. Indeed, when $\I (k_0\cos s)<0$, 
the functions $\GF_j^+(k_0\cos s)$ 
need to be continued analytically from the domain  $\I (k_0\cos s)>0$. Conversely, if
$\I (k_0\cos s)>0$, 
the functions $\GF_j^-(k_0\cos s)$ 
need to be continued analytically from the domain  $\I (k_0\cos s)<0$. This can be done by
the relations
$$
\GF^+(k_0\cos s)=\fr{1}{\Gd_0(k_0\cos s)\Gd_1(k_0\cos s)}G^*(k_0\cos s)\GF^-(k_0\cos s),
$$
\beq
\GF^-(k_0\cos s)=\fr{1}{\Gd_0(-k_0\cos s)\Gd_1(k_0\cos s)}G^{**}(k_0\cos s)\GF^+(k_0\cos s),\label{6.21}
\eeq
where
$$
G^*(k_0\cos s)=\left(\begin{array}{cc}
g_{11}^* & g_{12}^*\\
g_{21}^* & g_{22}^*\\
\end{array}
\right), \quad G^{**}(k_0\cos s)=\left(\begin{array}{cc}
g_{22}^* & -g_{12}^*\\
-g_{21}^* & g_{11}^*\\
\end{array}
\right),
$$$$
\Gd_0(k_0\cos s)=-(\Gg_1^--k_0\cos s)(\Gg_4^--k_0\cos s)-k_0^2\cos^2\Gb\sin^2 s,
$$
\beq
\Gd_1(k_0\cos s)=(\Gg_1^++k_0\sin s)(\Gg_4^++k_0\sin s)+k_0^2\cos^2\Gb\cos^2 s,
\label{6.22}
\eeq
and  $g^*_{sj}$ are entire functions expressed through polynomials of $\cos s$ and $\sin s$.
The same argument is applied to the functions  $\GF_j^\pm(k_0\sin s)$.
Since the surface waves do not affect the constants $C_j$, we do not specify
the surface waves poles.

Before we write the analytic continuation of the spectra introduce the integrals
$$
I_{j\pm}^c(s,a)=-\fr{1}{2\pi i}\int_{-i\infty}^{i\infty}F_{j\pm}(\Gs+a)\cot(\Gs-s)d\Gs,\quad a<\R s<\pi+a,
$$
\beq
I_{j\pm}^t(s,a)=-\fr{1}{2\pi i}\int_{-i\infty}^{i\infty}F_{j\pm}(\Gs+a)\tan(\Gs-s)d\Gs,\quad -\pi/2+a<\R s<\pi/2+a.
\label{6.20.1}
\eeq
We come finally to the analytic continuation of the functions $S_1(s)$ and $S_2(s)$
to the left and to the right that we need to recover the incident and reflected waves
$$
S_j(s)=I_{j-}^c(s,0)+I_{j+}^t(s,0)+i_j\cot(s-\Gt_0)-i_j\cot(s+\Gt_0)], \quad 0<\R s<\fr{\pi}{2},
$$$$
S_j(s)=I_{1-}^c(s,0)-I_{j+}^c(s,\pi/2)+i_j\cot(s-\Gt_0)
-(i_j-\GL_{j+})\cot(s+\Gt_0)
$$$$
+S_j^{s}(s;\pi/2,\pi), \quad \fr{\pi}{2}<\R s<\pi,
$$
$$
S_j(s)=-I_{j-}^t(s,\pi/2)+I_{j+}^t(s,\pi)+(i_j-\GL_{j-}+M_{j+})\cot(s-\Gt_0)
$$$$
-(i_j-\GL_{j+})\cot(s+\Gt_0)+S_j^{s}(s;\pi,3\pi/2),
\quad \pi<\R s<\fr{3\pi}{2},
$$
$$
S_j(s)=-I_{j-}^t(s,-\pi/2)+I_{j+}^t(s,0)+i_j\cot(s-\Gt_0)
$$
$$
-(i_j-\GL_{j-})\cot(s+\Gt_0)+S_j^{s}(s;-\pi/2,0), \quad -\fr{\pi}{2}<\R s<0,
$$
$$
S_j(s)=I_{j-}^c(s,-\pi)-I_{j+}^c(s,-\pi/2)+(i_j-\GL_{j+}+M_{j-})\cot(s-\Gt_0)
$$
\beq 
-(i_j-\GL_{j-})\cot(s+\Gt_0)+S_j^{s}(s;-\pi,-\pi/2),\quad 
 -\pi<\R s<-\fr{\pi}{2},
\label{6.20.2}
\eeq
where $S_j^{s}(s;a,b)$ are meromorphic functions whose
poles in the strip $a<\R s<b$ may generate the surface waves.

\subsection{Additional physical conditions}

Let $\CC$ be a closed contour that comprises the Sommerfeld contours $\CT_+$ and $\CT_-$ and two contours
$\CG_L$ and $\CG_R$.  The starting points of the contours $\CG_L$ and $\CG_R$ are $-3\pi/2+i\infty$ and
$3\pi/2-i\infty$, and they pass through the points $s=-\pi$ and $s=\pi$, respectively. 
The contour  $\CG_R$ is described by the equation  $\R s=\pi-{\rm gd}(\I s)\sgn(\I s)$, and the contour
$\CG_L$ is symmetric to $\CG_R$ with respect to the origin. 
Here, ${\rm gd}\, x$ 
is the Gudermann function
${\rm gd}\, x=\cos^{-1}{\rm sech}\, x$.
As $k_0r\to \infty$, the electric and magnetic field can be represented as
$$
E_z\sim
E_z^i+E_{z+}^r+E_{z-}^r+E_{z+}^{R}+E_{z-}^{R}+E_z^s+E_z^d,
$$
\beq
H_z\sim
H_z^i+H_{z+}^r+H_{z-}^r+H_{z+}^{R}+H_{z-}^{R}+H_z^s+H_z^d,\quad k_0r\to\infty,
\label{3.6''}
\eeq
$E_z^s$, $H_z^s$ are the surface waves, and $E_z^d$, $H_z^d$ are the diffracted waves.
Write down the poles of the functions $S_1(s)$ and $S_2(s)$  in the interval
$(-\pi,\pi)$ which generate the incident and reflected waves. These poles are
$$
 s=-\Gt+\Gt_0\in (-\pi/2,\pi/2),\quad s+\Gt\in(0,\pi/2),
$$
$$
 s=\pi-\Gt-\Gt_0\in (0,\pi), \quad s+\Gt\in(\pi/2,\pi),
$$
$$
 s=-\Gt-\Gt_0\in (-\pi,0), \quad s+\Gt\in(-\pi/2,0),
$$
$$
 s=\pi-\Gt+\Gt_0\in (\pi/2,\pi) \;\; {\rm if}\;\; \Gt_0<\Gt<\pi/2, \quad s+\Gt\in(\pi,3\pi/2),
$$
\beq
 s=-\pi-\Gt+\Gt_0\in (-\pi,-\pi/2) \;\; {\rm if}\;\; 0<\Gt<\Gt_0, \quad s+\Gt\in(-\pi,-\pi/2).
\label{6.28}
\eeq
The incident and reflected electrical and magnetic waves can be recovered from (\ref{6.1}) and (\ref{6.20.2}) 
$$
 E_z^i= i_1e^{-ik_0\Gr\cos(\Gt-\Gt_0)},\quad  ZH_z^i= i_2e^{-ik_0\Gr\cos(\Gt-\Gt_0)},
$$
$$
E_{z+}^r=(-i_1+\GL_{1+})e^{ik_0\Gr\cos(\Gt+\Gt_0)},\quad 
ZH_{z+}^r=(-i_2+\GL_{2+})e^{ik_0\Gr\cos(\Gt+\Gt_0)}, 
$$
$$
E_{z-}^r=(-i_1+\GL_{1-})e^{-ik_0\Gr\cos(\Gt+\Gt_0)},\quad 
ZH_{z-}^r=(-i_2+\GL_{2-})e^{-ik_0\Gr\cos(\Gt+\Gt_0)}, 
$$
$$
E_{z+}^R=(i_1+M_{1-}-\GL_{1+})e^{ik_0\Gr\cos(\Gt-\Gt_0)}\Go(\Gt;0,\Gt_0),
$$$$
ZH_{z+}^R=(i_2+M_{2-}-\GL_{2+})e^{ik_0\Gr\cos(\Gt-\Gt_0)}\Go(\Gt;0,\Gt_0),
$$
$$
E_{z-}^R=(i_1+M_{1+}-\GL_{1-})e^{ik_0\Gr\cos(\Gt-\Gt_0)}\Go(\Gt;\Gt_0,\pi/2),
$$\beq
ZH_{z-}^R=(i_2+M_{2+}-\GL_{2-})e^{ik_0\Gr\cos(\Gt-\Gt_0)}\Go(\Gt;\Gt_0,\pi/2).
\label{6.29}
\eeq 
The diffracted waves are determined in a standard manner in the form
\beq
E_z^d=\fr{e^{ik_0\Gr}}{\sqrt{k_0\Gr}}D_1(\Gt), \quad ZH_z^d=\fr{e^{ik_0\Gr}}{\sqrt{k_0\Gr}}D_2(\Gt),
\label{6.34}
\eeq
where
$D_1(\Gt)$ and $D_2(\Gt)$ are the diffraction coefficients
\beq
D_j(\Gt)=\fr{e^{-i\pi/4}}
{\sqrt{2\pi}}
[S_j(\Gt-\pi)-S_j(\Gt+\pi)]. \quad j=1,2.
\label{6.35}
\eeq
As for the surface waves, they can be recovered in the same way as the incident and reflected waves as soon as the location of the surface wave poles
is determined.

Now, the reflected coefficients in $(\ref{6.29})$  have to be consistent with those
given by (\ref{3.7.1}). This immediately determines the unknown constants $C_1$ and $C_2$
\beq
C_1=r_1^++i_1, \quad C_2=r_2^++i_2.
\label{6.36}
\eeq
As for the other six relations
\beq
\GL_{j-}=r_j+i_j, \quad M_{j-}=r_j^++R_j^+, \quad M_{j+}=r_j^-+R_j^-, \quad j=1,2,
\label{6.37} 
\eeq
they are satisfied identically.
On verifying these equalities
we may test the validity of the computations. Notice that the coefficients $\GL_{j-}$, $M_{j-}$ and $M_{j+}$
depend upon the entries of the matrix $G(\Gn_0)$ 
and do not depend on the actual solution to the vector RHP.

\vspace{.1in}

\setcounter{equation}{0}

\section{Normal incidence}

Consider the scalar case when $\Ga=\Gb=\pi/2$. The matrix coefficient $G(\Gn)$ of both vector RHPs,  1 and 2,  is a diagonal  matrix,
$$
G(\Gn)=\diag\{(\Gg_1^-+\Gn)(\Gg_1^--\Gn)^{-1}, (\Gg_4^-+\Gn)(\Gg_4^--\Gn)^{-1} \},
$$
\beq
\hat G(\Gn)=\diag\{(\Gg_1^++\Gn)(\Gg_1^+-\Gn)^{-1}, (\Gg_4^++\Gn)(\Gg_4^+-\Gn)^{-1} \},
\label{s1}
\eeq
and the boundary conditions (\ref{3.6}) and (\ref{3.6'}) become
\beq
\GF^+_j(\Gn)=\fr{\Gg_{j-}+\Gn}{\Gg_{j-}-\Gn}\GF_j^-(\Gn), \quad \Gn\in L, \quad j=1,2,
\label{s2}
\eeq
and 
\beq
\hat\GF^+_j(\Gn)=\fr{\Gg_{j+}+\Gn}{\Gg_{j+}-\Gn}\hat\GF_j^-(\Gn), \quad \Gn\in L, \quad j=1,2,
\label{s2'}
\eeq
respectively. Here  we adopted the notations $\Gg_{1\pm}=\Gg_{1}^\pm$ and  $\Gg_{2\pm}=\Gg_{4}^\pm$.
The coefficients
of the problems are rational functions, and the solution can easily be derived. 
 The polynomials $\Gd_0(\Gn)$ and $\hat\Gd(\Gn)$ have the form $\Gd_0(\Gn)=-(\Gn-\Gg_{1-})(\Gn-\Gg_{2-})$
 and $\hat\Gd_0(\Gn)=-(\Gn-\Gg_{1+})(\Gn-\Gg_{2+})$.
 As before, we consider the cases (i), (ii) and (iii).
 In the first case, both zeros of the polynomials $\Gd_0(\Gn)$ and $\hat\Gd_0(\Gn)$ lie in the lower half-plane,  
 $\I\Gg_{j-}<0$,   $\I\Gg_{j+}<0$, $j=1,2$,  and the solution to each RHP in (\ref{s2}) has two arbitrary constants, $D_{0j}$ and $D_{1j}$, 
 \beq
\GF_j^\pm(\Gn)=\fr{D_{0j}+D_{1j}\Gn^2}{(\Gn^2-\Gn_0^2)(\Gg_{j-}\mp\Gn)},\quad 
\Gn\in{\Bbb C^\pm}, \quad j=1,2.
 \label{s3}
 \eeq
 On employing (\ref{3.2}) we can derive a formula for $\hat\Gf_j(i\Gz,0)$
\beq
\hat\Gf_j(i\Gz,0)=-\fr{(\Gg_{j+}+i\Gz)(D_{0j}+D_{1j}\Gn^2)}
{(\Gn^2-\Gn_0^2)(\Gg_{j-}^2-\Gn^2)}.
\label{s4}
\eeq
We next denote $i\Gz=\hat\Gn$ and have
\beq
\hat\Gf_j(\pm\hat\Gn,0)=\fr{(\Gg_{j+}\pm\hat\Gn)[D_{0j}+D_{1j}(k_0^2-\hat\Gn^2)]}{(\hat\Gn^2-\hat\Gn_0^2)(\Gg_{j-}^2-k_0^2+\hat\Gn^2)},\quad 
\hat\Gn\in{\Bbb C^\pm}, \quad j=1,2.
\label{s5}
\eeq
On the other hand, the decoupled RHP 2 (\ref{s2'}) yields an alternative formula for the functions $\hat\Gf_j(\pm\Gn,0)=\hat\GF_j^\pm(\Gn)$,
\beq
\hat \GF_j^\pm(\Gn)=\fr{\hat D_{0j}+\hat D_{1j}\Gn^2}{(\Gn^2-\hat\Gn_0^2)(\Gg_{j+}\mp\Gn)},\quad 
\Gn\in{\Bbb C^\pm}, \quad j=1,2,
 \label{s6}
 \eeq
where  $\hat D_{0j}$ and $\hat D_{1j}$ are arbitrary constants. The compatibility condition 
\beq
\hat\Gf_j(\pm\Gn,0)|_{RHP\, 1}=\hat\GF_j^\pm(\Gn)|_{RHP\, 2}
\label{s6'}
\eeq
reads
\beq
(\Gg_{j+}^2-\Gn^2)[D_{0j}+D_{1j}(k_0^2-\Gn^2)]=(\hat D_{0j}+\hat D_{1j}\Gn^2)(\Gg_{j-}^2-k_0^2+\Gn^2), \quad \Gn\in{\Bbb C}.
\label{s7}
\eeq
These six conditions for the eight constants are fulfilled if and only if
\beq
D_{1j}=\hat D_{1j}=-\fr{D_{0j}}{\Gg_{j-}^2}, \quad \hat D_{0j}=\fr{\Gg_{j+}^2D_{0j}}{\Gg_{j-}^2}.
\label{s8}
\eeq
In this case the solution of both RHPs 1 and 2 has one arbitrary constant, $D_j=D_{0j}/\Gg_{j-}^2$,
\beq
\GF^\pm(\Gn)=\fr{D_{j}(\Gg_{j-}\pm\Gn)}{\Gn^2-\Gn_0^2},\quad 
\hat\GF^\pm(\Gn)=\fr{D_{j}(\Gg_{j+}\pm\Gn)}{\Gn^2-\hat\Gn_0^2},\quad 
\Gn\in{\Bbb C^\pm}, \quad j=1,2.
 \label{s9}
 \eeq

Consider next the second case  when $\I\Gg_{j-}<0$ and $\I\Gg_{j+}>0$, $j=1,2.$ 
As before, the functions $\hat\Gf_j(\pm i\Gz,0)=\hat\GF^\pm_{j}(\tilde\Gn)$ 
derived from the solution
of the RHP 1,  have the form
(\ref{s4}).  The same functions $\hat\GF^\pm(\Gn)$ comprise the solution to the RHP 2 and coincide with (\ref{s9}). The formulas (\ref{s5})
and (\ref{s9}) are  compatible 
 if and only if 
 \beq
 D_{0j}+D_{1j}(k_0^2-\Gn^2)=\hat D_j(\Gg_{j-}^2-k_0+\Gn^2).
 \label{s9'}
 \eeq
These four equations for six constants give. 
$D_{0j}=\Gg_{j-}^2D_j$ and $D_{1j}=-D_j$. Then the functions $\GF^\pm(\Gn)$ are the same as the ones in (\ref{s9}).

In the last case, $\I\Gg_{j-}>0$ and $\I\Gg_{j+}>0$, $j=1,2,$ by solving the RHPs 1 and 2 we find that the compatibility condition (\ref{s6'}) gives $D_j=\hat D_j$, and  the functions $\GF_j^\pm(\Gn)$
and $\hat\GF_j^\pm(\Gn)$ coincide with the functions given in (\ref{s9}).

Our intention next is to fix the constants $D_j$.
In all the three cases, regardless of whether $\I\Gg_{j\pm}$ is positive or negative, the functions $\GF_j^\pm(\Gn)$  have the same form (\ref{s9}).
Formulas (\ref{6.13}) and (\ref{s9}) imply
\beq
C_j=-\fr{iD_j(\Gg_{j-}+\Gn_0)}{2\Gn_0},
\label{s10}
\eeq
and from (\ref{6.36}) we fix the constants $D_j$
\beq
D_j=\fr{4i\Gn_0\hat\Gn_0 i_j}{(\Gn_0+\Gg_{j-})(\hat\Gn_0+\Gg_{j+})}, \quad j=1,2.
\label{s11}
\eeq
On putting in (\ref{6.19}) $\Gb=\pi/2$ we obtain
$$
\mu_{12}=\mu_{21}=0,\quad \mu_{jj}=\fr{\Gg_{j-}-\Gn_0}{\Gg_{j-}+\Gn_0}, \quad \GL_{j-}=\fr{2\Gn_0 i_j}{\Gg_{j-}+\Gn_0},
$$
\beq
M_{j-}=\fr{2\Gn_0(\hat\Gn_0-\Gg_{j+})i_j}{(\Gn_0+\Gg_{j-})(\hat\Gn_0+\Gg_{j+})}, \quad 
M_{j+}=\fr{2\hat\Gn_0(\Gn_0-\Gg_{j-})i_j}{(\Gn_0+\Gg_{j-})(\hat\Gn_0+\Gg_{j+})}.
\label{s12}
\eeq
Employing (\ref{3.7.1}) and (\ref{3.7.2}) we reduce the formulas for the reflection coefficients 
\beq
r^+_j=\fr{\hat\Gn_0-\Gg_{j+}}{\hat\Gn_0+\Gg_{j+}}i_j, \quad 
r^-_j=\fr{\Gn_0-\Gg_{j-}}{\Gn_0+\Gg_{j-}}i_j,
\quad
R_j^+=R_j^-=\fr{(\hat\Gn_0-\Gg_{j+})(\Gn_0-\Gg_{j-})}{(\hat\Gn_0+\Gg_{j+})(\Gn_0+\Gg_{j-})}i_j.
\label{s13}
\eeq
Simple derivations show that the six relations (\ref{6.37}) are identities.

\section*{Conclusions}

We have found a closed-form solution of the classical problem on scattering of an electromagnetic plane wave obliquely incident from an impedance right-angled concave
wedge.
Two Helmholtz equations coupled by the boundary conditions have been reduced  to two vector RHPs subject to the symmetry
condition for the unknown vectors, $\GF^+(\Gn)=\GF^-(-\Gn)$ and $\hat\GF^+(\Gn)=\hat\GF^-(-\Gn)$.  The unknown vectors $\GF^+(\Gn)$
and $\hat\GF^+(\Gn)$  are the Laplace transform of the electric and magnetic components on the vertical and horizontal faces of the wedge,
respectively. It has been found convenient not to specify 
the residues of the solution to the RHPs
at the geometric optics poles  at this stage. 
The matrix factorization problem has been solved by the 
technique of the RHP on a genus-3 Riemann surface in the case of the boundary conditions (\ref{1.1}).
In the general case  (\ref{1.2}) of the boundary conditions the matrix can be factorized as well. However, this requires
tedious computations of the analogues of formulas (\ref{3.13}), (\ref{3.16}) and (\ref{3.16'}).
Due to the symmetry of the problem we have managed to avoid the genus-3 Riemann $\Gt$-function and  solved the Jacobi inversion problem 
in terms of elliptic functions. The Wiener-Hopf matrix-factors have been found by quadratures and, eventually,
an exact solution to the two vector RHPs 
 associated with the diffraction problem of interest has been constructed. 
Either vector RHP is unconditionally solvable and has $\Gk+3$ (RHP 1) and $\hat\Gk+3$ (RHP 2) free constants,
where $2\Gk=\ind \det G(\Gn)$,  $2\hat\Gk=\ind \det \hat G(\Gn)$, and $G$ and $\hat G$ are the matrix coefficients of the vector RHP 1 and 2.
Depending on the location of the zeros of quadratic polynomials associated with each RHP the integers
$\Gk$ and $\hat \Gk$ can be 1, 0 and -1. 
We have shown that the solutions to the RHPs are not independent and have to satisfy certain compatibility conditions.
When satisfied these conditions reduce the number $n$ of free constants to 2 and 
therefore make $n$ independent
of the indices $\Gk$ and $\hat\Gk$. 
The electric and magnetic field components $E_z$ and $H_z$ have 
been derived in terms of  Sommerfeld's integrals with the spectra expressed through 
the  solution of the vector RHP 1.
By analytic continuation of the spectra to the left and to the right we have determined the poles
responsible for the reflection waves $E_{z+}^r$ and $H_{z+}^r$
and  fixed  the two free constants left.
We have also considered the case of normal incidence, derived a closed-form solution to the RHPs 1 and 2,
verified the compatibility conditions and recovered the incident and reflected waves.

Although we have not implemented numerical computations, the core of
the procedure, the factorization method on a Riemann surface, has been tested several times 
({\bf(\ref{antmoi})} and {\bf(\ref{antsil2})} in the genus-1 case  and {\bf(\ref{ant1})}
in the genus-3 case). Certainly, the technique is robust, and the determination  of the diffraction coefficients 
in (\ref{6.35})
for example requires computing just 
several integrals given in (\ref{6.12}), (\ref{3.58}) and (\ref{3.51}).

\vspace{.1in}

{\centerline{\Large\bf  References}}

\begin{enumerate}

\item\label{mal1}
G. D. Maliuzhinets, Excitation, reflection and emission of surface waves from a wedge with
given face impedances, {\it Sov. Phys. Dokl.} {\bf 3} (1958) 752-755.

\item\label{hur} R. A. Hurd and E. L¨uneburg, Diffraction by an anisotropic impedance half plane, {\it Canad. J.
Phys.} {\bf 63} (1985) 1135-1140.

\item\label{lue} E. L\"uneburg and A. H. Serbest, Diffraction of an obliquely incident plane wave by a two-face
impedance half plane: Wiener-Hopf approach, {\it Radio Sci.} {\bf 35} (2000) 1361-1374.

\item\label{ant1} 
Y. A. Antipov and V. V.  Silvestrov,  
Electromagnetic scattering from an anisotropic impedance half plane
at oblique incidence: the exact solution, {\it  Quart. J. Mech. Appl. Math.} {\bf 59} (2006) 211-251.

\item\label{ber}
J. M. L. Bernard, Diffraction at skew incidence by an anisotropic wedge in electromagnetism theory: a new class of canonical cases, {\it J. Phys. A: Math. Gen.} {\bf 31} (1998) 595-613.

\item\label{lya1} 
M. A. Lyalinov and N. Y. Zhu, Exact solution to diffraction problem by wedges with a class of anisotropic 
impedance faces: oblique incidence of a plane electromagnetic wave, {\it IEEE Trans. Antenn. Propag.} {\bf 51}
(2003) 1216-1220.

\item\label{sye}
H. H. Syed and J. L. Volakis,
An approximate solution for scattering by an impedance wedge
at skew incidence, {\it Radio Sci.} {\bf 30} (1995) 505-524.

\item\label{pel} G. Pelosi, G. Manara and P. Nepa, 
A UTD solution for the scattering by a wedge with anisotropic impedance faces: skew incidence case 
{\it IEEE Trans. Antenn. Propag.} {\bf 46} (1998) 579-588. 

\item\label{lya2} 
M. A. Lyalinov and N. Y. Zhu, Diffraction of a skew incident plane electromagnetic wave
by an impedance wedge, {\it Wave Motion} {\bf 44} (2006) 21-43.

\item\label{ant2} 
Y. A. Antipov and V. V.  Silvestrov, 
Method of integral equations for systems of difference equations, {\it IMA J. Appl. Math.} {\bf 72} (2007) 681-705.

\item\label{lya} M. A. Lyalinov and N. Y. Zhu, {\it Scattering of waves by wedges and cones with impedance boundary conditions} (SciTech Publ. Inc., Raleigh, 2012).

\item\label{dan1} V. G. Daniele and G. Lombardi,  
Wiener-Hopf solution for impenetrable wedges at skew Incidence, 
{\it IEEE Trans. Antenn. Propag.} {\bf 54} (2006) 2472-2485. 

\item\label{bud}
B. Budaev and D. B. Bogy, 
Diffraction of a plane skew electromagnetic wave by a wedge with general anisotropic impedance boundary conditions, 
{\it IEEE Trans. Antenn. Propag.} {\bf 54} (2006) 1559-1567. 

\item\label{sha} A. V. Shanin, On wave excitation in a wedge-shaped region, {\it Acoustical physics}
{\bf 42} (1996) 696-701.

\item\label{osi1} A. V. Osipov and T. B. A. Senior, 
Diffraction and reflection of a plane electromagnetic wave by a right-angled impedance wedge, 
{\it IEEE Trans. Antenn. Propag.} {\bf 57} (2009) 1789-1803. 

\item\label{osi2} A. V. Osipov and T.B.A. Senior, Electromagnetic diffraction by arbitrary-angle impedance
wedges, {\it Proc. Roy. Soc. A} {\bf 464} (2008) 177-195.

\item\label{dan2} V. G. Daniele, The Wiener-Hopf technique for impenetrable wedges having arbitrary aperture
angle, {\it SIAM J. Appl. Math.} {\bf 63} (2003) 1442-1460.

\item\label{ant4} Y. A. Antipov, 
A symmetric Riemann-Hilbert problem for order-4 vectors in diffraction theory, 
{\it Quart. J. Mech. Appl. Math.} {\bf 63} (2010) 349-374. 

\item\label{ant5} Y. A. Antipov, A genus-3 Riemann-Hilbert problem and diffraction of a wave by orthogonal resistive half-planes, {\it Comput. Meth.  Function Theory} {\bf 11} (2011) 439-462.

\item\label{moi} N. G. Moiseev, Factorization of matrix functions of special form, {\it Sov. Math. Dokl.}
{\bf 39} (1989)
264-267.

\item\label{antmoi}
Y.A. Antipov and N. G. Moiseev, Exact solution of the plane problem for a composite plane with a cut across the boundary between two media, {\it J. Appl. Math. Mech. (PMM)}
{\bf 55}
(1991) 531-539.

\item\label{antsil}
Y.A. Antipov and V.V. Silvestrov, Factorization on a Riemann surface in scattering theory, {\it Quart. J. Mech. Appl. Math.}  {\bf 55} (2002) 607-654.

\item\label{zve} E. I. Zverovich, Boundary value problems in the theory of analytic functions in H\"older classes on Riemann surfaces, {\it Russian Math. Surveys} {\bf 26} (1971) 117-192.

\item\label{han} H. Hancock, {\it Lectures on the Theory of Elliptic Functions} (Dover, New York 1968).

\item\label{mal2}
G. D. Maliuzhinets, Relation between the inversion formulas for the Sommerfeld integral and the formulas of Kontorovich-Lebedev,
{\it Sov. Phys. Dokl.} {\bf 3} (1958) 266-268.

\item\label{antsil2} Y.A. Antipov and V.V. Silvestrov, Second-order functional-difference equations. II.: Scattering from a right-angled conductive wedge for E-polarization, {\it Quart. J. Mech. Appl. Math.} 
 {\bf 57} (2004) 267-313.

\end{enumerate}

\end{document}